# Analyzing Big Data with Dynamic Quantum Clustering

M. Weinstein,[1][*][+], F. Meirer[2], A. Hume[3], Ph. Sciau[4], G. Shaked[5], R. Hofstetter[6], E. Persi[5], A. Mehta[1], and D. Horn[5].

* Corresponding author . email: niv@slac.stanford.edu
[1] SLAC National Accelerator Laboratory, 2575 Sand Hill Rd., Menlo Park, CA 94025
[2] Inorganic Chemistry and Catalysis, Utrecht University, Universiteitsweg 99, 3584 CG Utrecht, Netherlands
[3] Dept. of Physics, Princeton University, Princeton, NJ
[4] CNRS, CEMES, BP 94347, 29 rue Jeanne Marvig, 31055 Toulouse, France
[5] School of Physics and Astronomy, Tel Aviv University, Tel Aviv 69978, Israel
[6] Geophysical Institute of Israel, Lod 71100, Israel

† This work was supported by the U. S. DOE, Contract No.~DE-AC02-76SF00515.

## Abstract

*How does one search for a needle in a multi-dimensional haystack without knowing what a needle is and without knowing if there is one in the haystack?* This kind of problem requires a paradigm shift - away from hypothesis driven searches of the data - towards a methodology that *lets the data speak for itself.* Dynamic Quantum Clustering (DQC) is such a methodology. DQC is a powerful visual method that works with big, high-dimensional data. It exploits variations of the density of the data (in feature space) and unearths subsets of the data that exhibit correlations among all the measured variables. The outcome of a DQC analysis is a movie that shows how and why sets of data-points are eventually classified as members of simple clusters or as members of - what we call - *extended* structures. This allows DQC to be successfully used in a non-conventional exploratory mode where one searches data for unexpected information without the need to model the data. We show how this works for big, complex, real-world datasets that come from five distinct fields: *i.e.,* x-ray nano-chemistry, condensed matter, biology, seismology and finance. These studies show how DQC excels at uncovering unexpected, small - but meaningful - subsets of the data that contain important information. We also establish an important new result: namely, that big, complex datasets often contain interesting *structures* that will be missed by many conventional clustering techniques. Experience shows that these structures appear frequently enough that it is crucial to know they can exist, and that when they do, they encode important hidden information. In short, we not only demonstrate that DQC can be flexibly applied to datasets that present significantly different challenges, we also show how a simple analysis can be used to *look for the needle in the haystack, determine what it is, and find what this means.*

## 1. Introduction

Data of all kinds is being acquired and warehoused at a tremendous rate. An important challenge that faces producers of *big* - often poorly understood - data is how to search it for unexpected information. This challenge is often neglected in favor of looking for known structures, as in conventional classification problems. Creating a search strategy when one doesn't know what one is looking for and when one isn't sure there is information in the data might appear to be impossible, yet this is what Dynamic Quantum Clustering (DQC) [1] does. Although DQC can be loosely described as a density based clustering algorithm, it is much more than that. It is better described as a visual tool that can reveal subsets of large, complex data that exhibit simultaneous correlations among the many variables being measured. A DQC analysis begins with the creation of a movie wherein proxies of the data-points move from their initial position towards the nearest region of higher density. Hereafter this step will be referred to as the DQC evolution of the data. Correlated subsets are distinguished from one another depending on their final shape during or after DQC evolution (see section 2.1): extended shapes are referred to as *structures,* while the term *cluster* is reserved for subsets that collapse to a point. A DQC analysis results in a movie that visually reveals how and why the algorithm identifies and distinguishes between structures and clusters. This focus on all of the stages of the process distinguishes DQC from conventional clustering methods that focus on end results. In this way DQC gives an investigator a novel tool that allows better characterization of information hidden in the data before engaging in complex statistical analyses.

Because DQC is *data*-agnostic - in that it doesn't have to use domain specific knowledge to partition the data – it can be usefully applied to data coming from all fields of endeavor. Because DQC doesn't begin by assuming there are structures to be found, and because it has been proven not to find structures in random data and it makes no assumptions about the type or shape (topology) of structures that might be hidden in the data, it can be used to determine if one is collecting the right kind of information. In contrast to methods that partition data based upon discernible separations, such as the support vector machine technique, DQC exploits variations in the density of the data. Thus, it reveals structures with unusual topologies even in very dense datasets. Furthermore, DQC works well for high-dimensional data since the time spent in a DQC analysis only grows linearly with the dimension of the data. Finally, while DQC's greatest strength is that it allows one to visually explore high-dimensional complex data for unexpected structure, it can also be used to rapidly classify incoming data once a sufficiently large subset of data has been analyzed. In this way it can be used much in the same way as a neural net or tuned decision-tree. The quantum mechanical underpinnings of the DQC algorithm make it possible to deal with highly distributed data in completely parallel fashion, which allows it to scale to very large problems.

To demonstrate the validity of these statements we will discuss the application of DQC to several different real-world datasets. These datasets come from different areas of science and all are challenging for more familiar data-mining techniques. While only the first two sets are large, all are complex and present unique challenges. The topics we will cover are:

- **Nano-Chemistry (TXM-XANES)**: X-ray absorption spectroscopy is often used to analyze local chemistry. This example demonstrates that DQC can be used to study a set of spectroscopic data taken from a fragment of a Roman urn. The goal is to use this data to determine the different oxidation states of iron that are present, as well as their spatial distribution. This requires grouping approximately 700,000 raw spectra according to subtle differences in shape, all without making any assumptions about the substances that are present. This is a daunting problem and DQC succeeds remarkably well at this task. The most important part of this analysis is the demonstration, for the first time, of important extended filamentary structures hidden in this large, complex dataset. These structures are not simple clusters, but they are demonstrably meaningful and not artifacts of the algorithm. The ability to analyze such a large, complex and noisy dataset without bias and with sensitivity to small amounts of impurities is important to many real world problems.

- **Condensed Matter Experiments (LCLS pump-probe):** The second dataset contains x-ray scattering data from a pump-probe experiment carried out at the SLAC Linear Coherent Light Source (LCLS). This is another large, complex and noisy dataset. The experimental challenge is to analyze the data and show that the clean data carries information about the phonon band structure of the crystal allowing one to study non-equilibrium phonon dynamics. The specific data comes from an experiment where a germanium crystal is pumped by an infra-red laser and then the time dependent response of the crystal is recorded by probing it with an x-ray beam. The detector is a 512 x 512 pixel array and so the data consists of a 512 x 512 array of time dependent signals. The goal is to study this raw data and identify those pixels that exhibit clean signals from all of the noise; then to show that the clean data shows pixel to pixel correlations that are evidence for coherent oscillations induced by the pulse from the infra-red laser. As in the case of the x-ray absorption data, this dataset revealed unexpected filamentary structures, but with the twist that the extended structure, that comprised more than 95% of the data, turned out to be unwanted background. By removing this background and redoing the DQC analysis we reached all of our goals.

- **Protein Function:** The third dataset is very different in nature and poses another very difficult problem. This dataset consists of aligned amino-acid sequences for two functionally different aquaporin proteins. In this case the problem is to show that the linear sequences contained enough information to distinguish the two kinds of aquaporins from one another and to identify those locations on the sequence that most probably determine the differences in functionality. This task is very complex because of the large number of amino-acids along the sequence. DQC was used to produce a natural filtering process that made this task feasible and lead to successful results.

- **Geophysics:** The last dataset consists of twenty years of earthquake data for the Middle East. Conventionally earthquakes are labeled by magnitude, location and time of occurrence, however other characteristics such as, stress drop, faulting radius, and corner frequency are also extracted from the seismograms and cataloged. This data is taken from a catalog of such parameters for earthquakes in the Eastern Mediterranean Region (EMR). The goal is to meaningfully characterize these earthquakes in terms of their physical parameters alone, without reference to time or geographic information. We show that this can be done and then show that structures revealed by the DQC analysis are strongly correlated with the corresponding time and location information and have geophysical interpretation.
- **Finance:** The last dataset consists of ten years of S&P 500 data. We show that this data reveals surprising structure when the ~2800 days are considered to be the data entries with the daily prices of those 400 stocks that exist throughout the period are treated as *features*. The picture that emerges from the analysis is that there are market epochs characterized by differences in the way these 400 stocks are correlated with one another.

**Why is DQC different?**

Although DQC can be loosely described as a density based clustering technique, it is much more than that. The algorithm borrows familiar ideas from quantum mechanics to construct a quantum potential that provides an excellent proxy for the density of the data in multi-dimensional feature space. It then uses quantum evolution to efficiently move a proxy for each datum to the nearest local minimum of this potential. At the end of the evolution data point proxies – from here on simply referred to as data-points - that collect at a single, well-defined local minimum of the potential are called a *cluster;* data points that collect along an extended valley in the potential (created by a region of nearly constant density in the data) are called a *structure*. The most important result that emerges from most of these analyses is that many datasets reveal such *topologically non-trivial structures* that encode hidden information.

**1.2    What is different about the user interface?**

As already noted, DQC exhibits the individual steps of the quantum evolution as frames in a movie allowing one to see how and why structures form and to determine when the clustering process is complete. This visual presentation of results provides an intuitive way of attacking the difficult task of searching large datasets for simultaneous correlations among tens to hundreds of variables. DQC differs from many standard statistical methods that seek to determine how well a dataset matches a pre-existing parameterized model because it does not assume the existence of any such model. Thus, DQC allows one to explore large, complex datasets and find *unexpected* information.

**2. Explaining DQC**

**2.1    The Algorithm**

A DQC analysis begins with data that is presented as an $m \times n$ data matrix. Each data point is one of the $m$ rows of the matrix and is defined by the $n$-numbers that appear in that row, These $n$ numbers are referred to as *features* and the set of all possible sets of $n$-values that might appear in a row is referred to as the *feature space*. The process of creating a clustering algorithm using ideas borrowed from quantum mechanics starts with the creation of a potential function [2] that serves as a proxy for the density of data points. We do this as follows:

- First, define a function – referred to as a Parzen estimator [3, 4] - on the $n$-dimensional feature space. This function is constructed as a sum of Gaussian functions centered at each data point; *i.e.*, for $m$ data points, $\vec{x}_l$, we define $\varphi(\vec{x})$ to be

$$\varphi(\vec{x}) = \sum_{l=1}^{m} e^{-\frac{1}{2\sigma^2}(\vec{x}-\vec{x}_l)\cdot(\vec{x}-\vec{x}_l)} . \qquad (2.1.1)$$

- Next derive a potential function $V(\vec{x})$ defined over the same n-dimensional space. Since $\varphi(\vec{x})$ is a positive definite function we can define $V(\vec{x})$ as that function for which $\varphi(\vec{x})$ satisfies the time-independent Schrödinger equation

$$-\frac{1}{2\sigma^2}\nabla^2\varphi + V(\vec{x})\varphi = E\varphi = 0. \qquad (2.1.2)$$

  - NOTE: The value zero is chosen to simplify the mathematics and plays no important role. Clearly the energy, **E**, can always be set to zero by adding a constant to the potential. It is straightforward to solve the above equation for $V(\vec{x})$.

- The quantum potential is of interest for two main reasons: first, physical intuition tells us that the local minima of $V(\vec{x})$ will correspond to the local maxima of $\varphi(\vec{x})$ if the latter are well separated from one another; second $V(\vec{x})$ may have minima at points where $\varphi(\vec{x})$ exhibits no corresponding maxima. If – as it was originally envisaged – the Parzen estimator is meant to be a proxy for the density of the data, then DQC's quantum potential can be thought of as a unbiased way of contrast enhancing the Parzen function to better reveal structure in the data. An additional benefit of working with this contrast enhanced version of the Parzen estimator is that its features depend much less sensitively upon the choice of parameter **σ** that appears in Eq.2.1.1. To summarize, the quantum potential reduces DQC's sensitivity to an arbitrary parameter and – at the same time – enhances the visibility of hard to see structure. Fig.1 shows how this works for a simple example, where we see that the potential reveals clear structure when the Parzen function shows – at best - a hint of structure. It also shows how the structure of the potential remains when **σ** is varied by 20%, even though the structure in the Parzen function disappears completely.

- Using the Hamiltonian defined by this potential, evolve each Gaussian that is associated with a specific data point by multiplying it by the quantum time-evolution operator $e^{-i\,\delta t\,H}$ (where $\delta t$ is chosen to be small). NOTE: this operator is constructed in the subspace spanned by all of the Gaussians corresponding to the original data points (details of this process are given in ref.[1]).
- Next, compute the new location of the center of each evolved Gaussian (hereafter referred to as the evolution of the data-point).
- Iterate this procedure. Ehrenfest's theorem guarantees that for small time steps, the center of each Gaussian will follow Newton's laws of motion, where the force is given by the expectation value of the gradient of the potential function. In this way DQC evolution is the analog of gradient-descent in classical mechanics. The fact that we use quantum evolution rather than more familiar classical methods allows us to convert the computationally intensive problem of gradient descent in a multi-dimensional potential into an exercise in matrix multiplication. This greatly reduces the workload and allows parallel execution of the code in order to quickly deal with enormous sets of data.

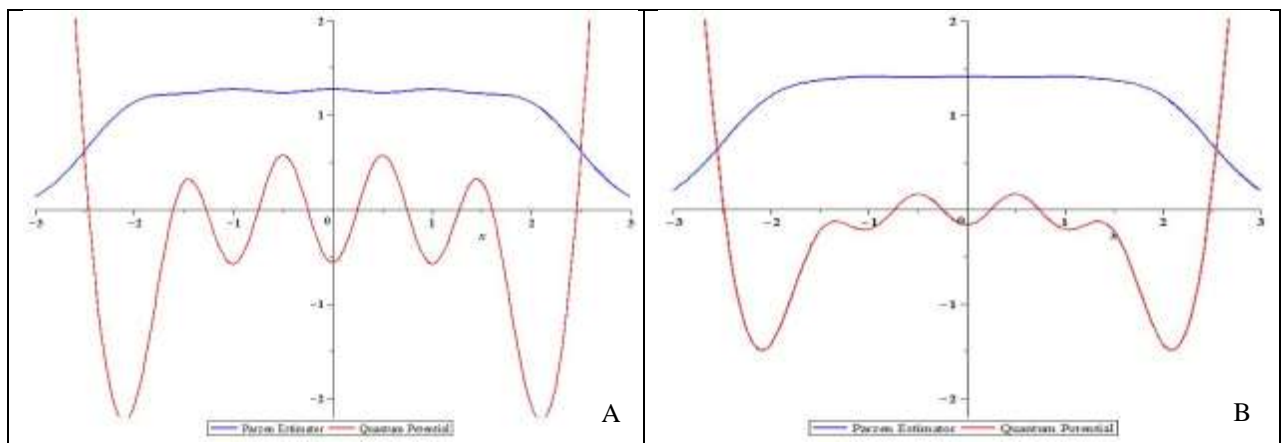

**Fig.1** Plots comparing the Parzen function to its corresponding quantum potential for values of $\sigma^2$ that differ by 20%. In each case the Parzen function is a sum of five Gaussian centered at the points -2,-1,0,1,2. **(A)** This plot corresponds to a smaller value of $\sigma^2$ and we see that the Parzen function shows barely discernible variations corresponding to location of the data points. Nevertheless, the corresponding quantum potential clearly exhibits the underlying structure. **(B)** This plot is for an 20% larger value of $\sigma^2$ and now it the variation of the Parzen function is essentially invisible, however the quantum potential still clearly shows the location of the data.

The output DQC evolution is an animation showing how data points move towards the nearest minimum of the potential. If the potential has isolated minima due to topologically simple regions of higher density, then the results of the evolution are fixed points describing isolated clusters. If, however, there are higher density regions of the data where the density is constant along complicated and possibly intersecting shapes, then the results of DQC evolution will be filamentary structures. This is what one will see if there are subsets of the data that exhibit multivariate correlations that can be parameterized in terms of only a few variables. Evolution of both the raw TXM-XANES and LCLS pump-probe data exhibit such filamentary regions. We will show that these structures encode important information about the data.

## 2.2 Sloan Digital Sky Survey Data

To demonstrate that the DQC potential accurately captures the density of data-points, and that DQC evolution can reveal extended, topologically non-trivial structures (or regions of nearly constant density) hidden in the data, we apply it to a well understood subset of 139,798 galaxies taken from the Sloan Digital Sky Survey (SDSS). Each data entry consists of the three coordinates of a single galaxy. The first two numbers are $\theta$ and $\varphi$, the angular coordinates defined in our Galaxy; the third coordinate is the redshift, $z$, a proxy for the distance from us to the other galaxies. It is well known that galaxies are not uniformly distributed, but rather they form a web of filaments and voids, so the question is if DQC evolution will reveal this structure.

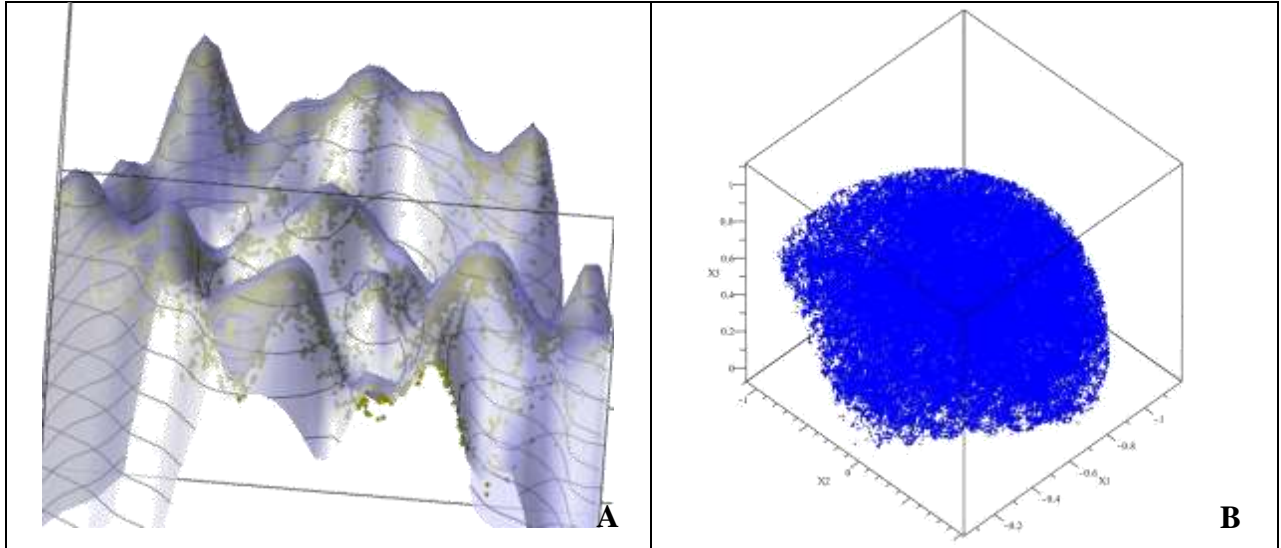

**Fig. 2A** Comparison of SDSS data points with the derived DQC potential. The potential is plotted upside down, and the yellow data points are slightly shifted in order to increase their visibility.

**Fig. 2B** The distribution of data in a 3D space defined by $\theta$ $\varphi$ and z.

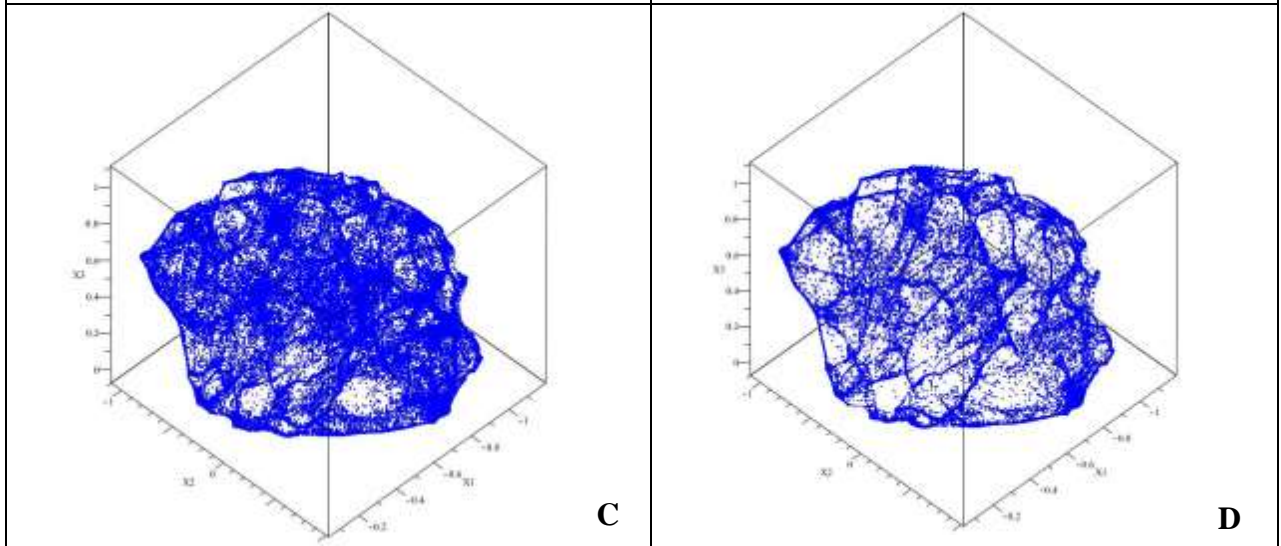

**Fig. 2C** Early stage of DQC evolution of the data

**Fig. 2D** Further DQC evolution exhibits the clear appearance of string-like structures.

Fig.2A is a plot of the quantum potential for a two-dimensional subset of the data obtained by choosing galaxies whose red-shifts differ by a very small amount. The galaxies are plotted as yellow points and the transparent quantum potential constructed from this set of galaxies is plotted upside-down, so that the maxima of the upside-down potential actually correspond to minima. This plot shows that the potential closely conforms to the distribution of galaxies and so it is clearly a very good proxy for the density of the data. As before, changing $\sigma$ by 20% doesn't change the potential significantly. Note that this two-dimensional slice of the data shows significant structure, but fails to exhibit filamentary features of nearly constant density. In Figs.2B-2D we see what happens to the full three-dimensional dataset as DQC evolution collects the data-points into structures that follow the shape of the minima of the three-dimensional potential. In this case DQC evolution reveals the existence of the network of filaments and voids that is not readily apparent in Fig.2B. The web of filaments revealed in this picture correspond to the topological structure of the minima of the quantum potential.

## 3. Analyzing X-ray Absorption Data

### 3.1    Background:
Materials that exhibit complex structure at many different scales abound in nature and technology. Examples of such materials are the electrodes of lithium-ion batteries, human bone and Roman pottery. Understanding these materials requires studying them with both very high resolution and a large field of view. This requires devices capable of collecting this kind of massive data and new methods to analyze this data once it has been collected.

X-ray absorption spectroscopy is used to study the chemical and physical structure of samples from biology, archeology, materials science, etc. The TXM-XANES microscope[5], located at the Stanford Synchrotron Radiation Lightsourc (SSRL), is a new device that enables researchers to efficiently study hierarchically complex materials[6]. In what follows we show how DQC successfully analyzes this kind of data without expert knowledge or *a-priori* assumptions.
The specific problem is the analysis of x-ray absorption data taken from a piece of glossy red and black Roman pottery dating back to the first century BCE. This large dataset, while small compared to what would result from the study of a functioning lithium-ion battery, is nevertheless a good proxy for the battery problem because the two problems share important common features. In particular, the oxidation-reduction chemistry occurring in these ceramics, due to percolation of oxygen through the hierarchy of nanometer to micron sized cracks and pores, is quite similar to the charging-discharging chemistry that occurs at the electrodes of a lithium-ion battery.

## 3.2 The Raw Data

The raw data consists of 669,559 x-ray absorption spectra[7]. Each x-ray absorption spectrum (XANES) records the absorption of x-rays by a single location on the sample. This absorption spectrum is measured at 146 distinct energies. Fig.3A displays 100 randomly chosen spectra of this type. The shape of any one spectrum contains information about the oxidation state of iron at one specific location. The goal is to produce a map of the sample showing the oxidation state of iron, the density of this iron oxide and the matrix of material other than iron that can be found in each 30nm x 30nm pixel. Identifying regions having the same composition amounts to identifying absorption spectra that have similar shapes. This kind of problem is usually studied by fitting the spectra to linear combinations of known reference spectra, but this approach suffers from serious ambiguities. DQC provides an unbiased way of approaching the problem.

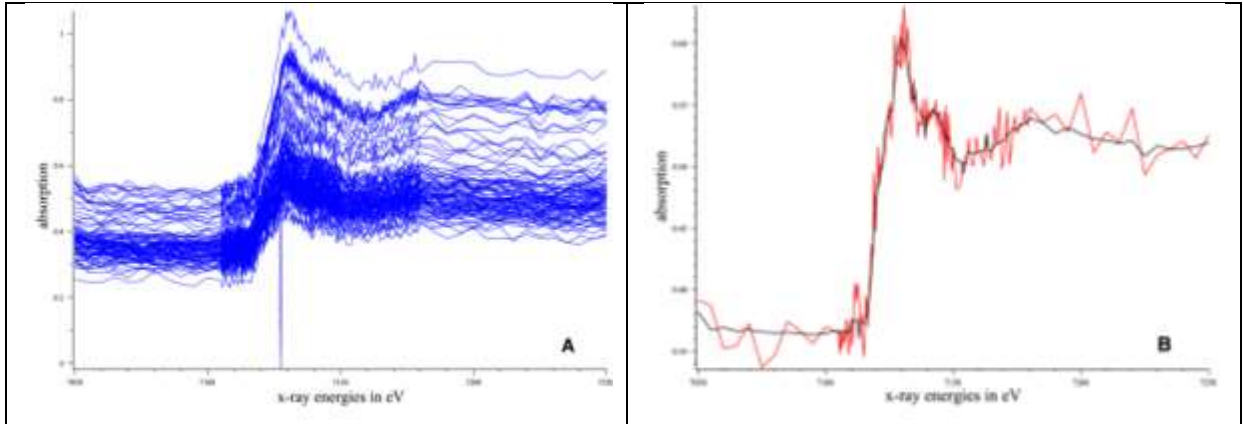

**Fig. 3** Comparison of raw and smoothed spectra. **(A)** Randomly chosen raw spectra that show differences in height of the pre-jump region (indicative of absorption due to the materials around the iron) and the edge-jump (differences in height between the pre-jump region and the height of the spectra above 7180 eV). The location of the beginning of the jump indicates different degrees of ionization for the iron atoms. **(B)** The red curve is a typical raw spectrum and the black curve shows the result of reducing the dimension of the data to the first five SVD components.

In order to rescale the data to fit into a unit cube in *n*-dimensions and in order to remove stochastic noise, we begin the analysis by constructing a Singular Value Decomposition (SVD) [4] for $M_{data}$, the 666,559 x 146 data matrix, each of whose rows is one absorption spectrum. SVD decomposes $M_{data}$ into a product of three matrices; *i.e.,*

$$M_{data} = U\,S\,V^{tr}.$$

$U$ is a 666,559 x 666,559 unitary matrix, $S$ is a 666,559 x 146 real matrix with entries only on the principal diagonal, and $V^{tr}$ is a 146 x 146 unitary matrix. The eigenvalues (ranked in decreasing magnitude) in $S$ fall off rapidly, and the slope of the fall-off changes markedly at the fifth eigenvalue.

We find that approximating $M_{data}$ by truncating $U$ to its first five columns, $S$ to its first five eigenvalues and $V^{tr}$ to its first five rows achieves a significant reduction in stochastic noise without losing the important features of most of the spectra. A typical example showing how well this works is presented in Fig.3B. While all of the higher SVD components contain mostly noise, we know that some subtle information is lost by removing them. Most important is that by approximating the current dataset by truncating to the first five SVD eigenvectors one partially suppresses a distinct spectral feature of a minority (~0.1%) component that spectroscopists use to identify un-oxidized iron. Despite this, we will demonstrate that DQC is still able to extract this minority phase from the 5-dimensional SVD representation. Note that, while we use the truncated SVD space for performing DQC clustering, we will only use averages of the raw data over clusters (or structures) to determine their chemical compositions.

### 3.3 Evolving the TXM-XANES Data

DQC evolution begins by assigning five-dimensional coordinates to each data point. This is done by associating a specific data entry with the corresponding row of $U$ (restricted to its first five columns and normalized to produce vectors of unit length). Fig. 4 shows a series of frames, the first of which is a plot of the starting configuration of the original data for dimensions 1-3. If one ignores the colors that are determined after evolution, it is clear that at the outset no obvious separation of the data exists in these coordinates. This lack of separation is typical of large complex datasets. The subsequent frames, going from left to right, top to bottom, show subsequent stages in the DQC evolution of the data. As is evident from these plots, we see that the blue points – shown in Fig.4A -separate and collapse, as shown in Fig.4B and become a single fixed-point by Fig.4D. This all occurs at a very early stage of the evolution. The rest of the data begins collapsing into strands, eventually stabilizing either as small clusters of fixed-points, or what can be described as string figures. As noted, the colors were assigned after the evolution was complete by selecting the clusters of fixed-points and string figures and coloring some of the parts of the final configurations for individual attention.

### 3.4 Drilling Down and Extracting the Local Chemistry

DQC evolution produced at least nine major topological features that are identified as extended structures. We used two approaches to decode the information encoded in these complex shapes. First, we searched for commonality among connected or adjacent structures. Second, we looked for variation and sub-structure within a shape. Following the first procedure, we averaged the raw spectra for those data points that belong to an individual component of the complex figures seen in the last plot of Fig. 4. For example, Fig.5A shows the averages of the raw spectra belonging to each of the four parts of the red structure ( the ``dancing man'' structure) in Fig. 4F. The average of each of the four substructures produces curves that look very similar, with most of the difference occurring in the absorption below the Fe edge (the sharp rise in the spectrum). Similarly, averages obtained from the green sub-structures also look very similar to one another. However, the green spectra are clearly significantly different from the red curves. We know that the absorption below the Fe edge is due to material that doesn't contain iron and we know that the difference in height between the absorption below the Fe edge and well above the Fe edge is related to the density of the iron oxide. This implies that if we remove this information from the raw data the curves should become identical. To accomplish this we implement a simple *normalization procedure* in which the average of the lowest 20 energy points is subtracted from the raw spectrum and the resulting spectrum is rescaled so the average of the highest 20 energy points is unity. Fig.5B shows that - as expected - computing the average for the components of the "dancing man" for the normalized data results in four curves that can't be distinguished from one another. Similarly, the same procedure applied to the green curves causes them to all collapse to a single curve. The different behaviors of the red and green average curves are quite similar to those of hercynite and hematite reference spectra; hence we refer to them by these labels.

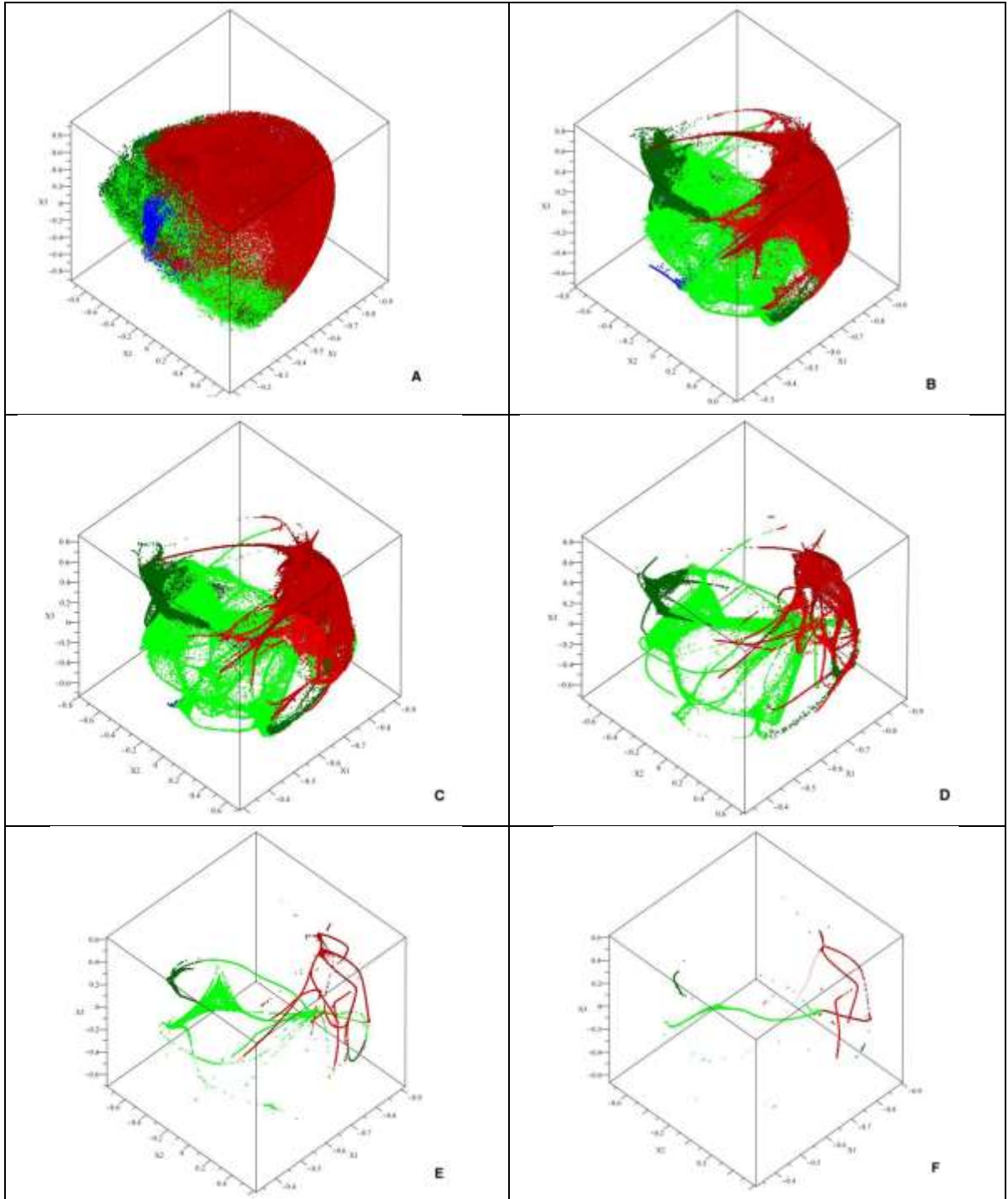

**Fig. 4** Reading from left to right, top to bottom we see the first three coordinates of the data plotted for frames 1, 5, 8, 15, 35, and 75 of the DQC animation. Colors were assigned to the fully evolved data, shown in the last frame. In frames **A**, **B** and **C,** we see the bluish points collapse to a very small cluster early in the evolution. Other points evolve slowly and don't collapse to a point, but rather form string-like figures. The string-like structures are stable over many frames. Similar structure is seen in dimensions 3, 4 and 5.

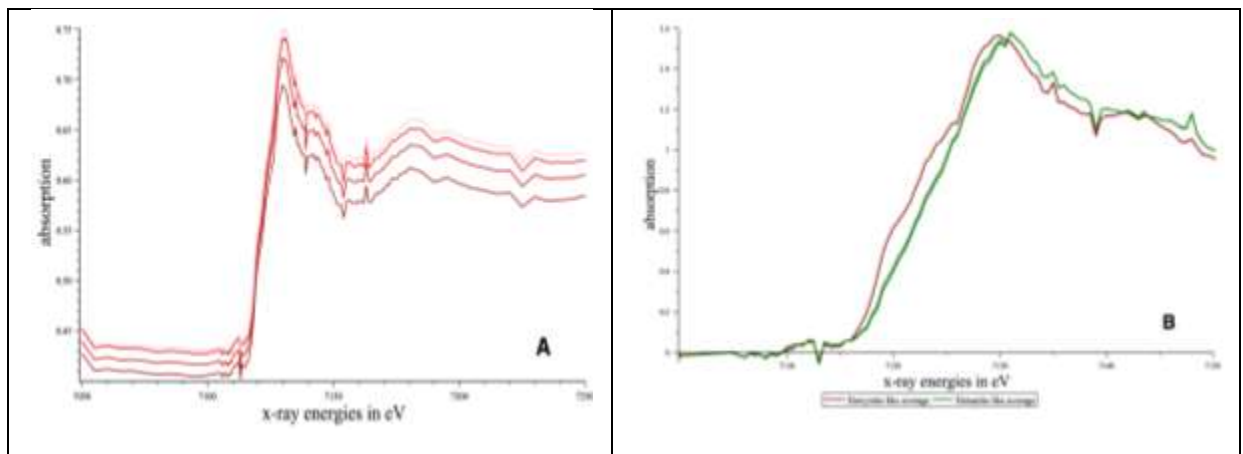

**Fig 5.** Plot of spectra associated to sections of the non-trivial topologies before and after normalization. **(A)** Each curve is a plot of the average of the raw absorption spectra for the different limbs of the ``dancing man'' structure shown in Fig.3F. The averages remove most of the noise from the raw data, so we see that although the spectra are very similar they differ in the value of the pre-edge and the edge-jump. **(B)** After normalization (see text) we see that the overlay of all of the averages for red structures (parts of the `` dancing man'') coincide, however the average of the limbs of the green structure is clearly different. The common spikes in the different curves are artifacts of the XANES data.

We next focus on the 2200 pixels that rapidly separated during the evolution of the full data (the blue area in Fig. 4A). We then compared the DQC evolution of the raw versus the normalized data for these points and found the evolution to be very similar. In each case a subset of 60 points immediately separated from the rest. Inspecting the raw spectra for these 60 points we see that some of these spectra closely match the spectrum of metallic Fe (see Fig.6A); whereas some of the others match magnetite (Fig.6B).

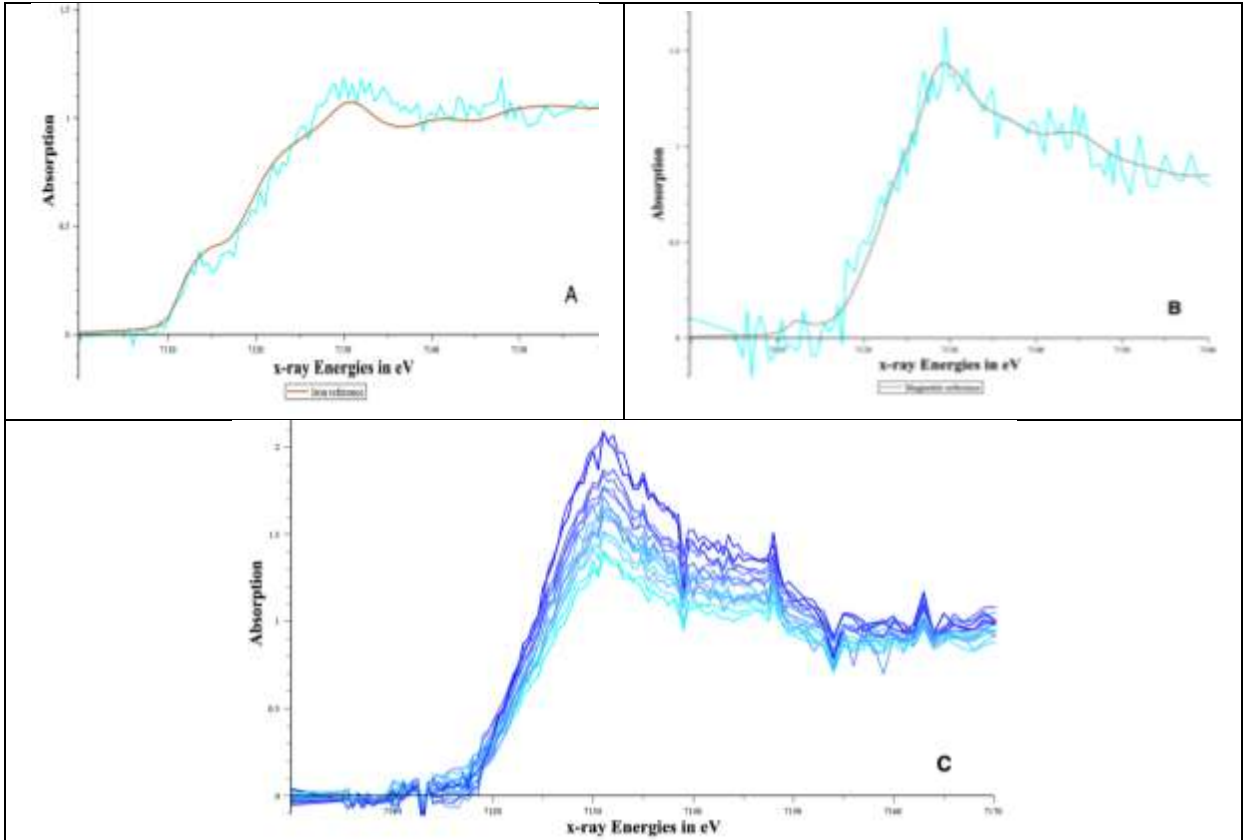

**Fig. 6.** Comparison of the individual raw spectra – shown in cyan- to reference spectra – shown in brown. **(A)** A reference iron spectrum compared to the spectrum for one of the 60 points contained in our *iron*-like cluster (see inset of Fig.6); **(B)** A reference magnetite spectrum to one of our *magnetite*-like spectra; **(C)** A plot of the average spectra for sub-clusters found in the bluish region, showing the range of variation of the averages.

### 3.5   Putting It All Together:

Finally, we assign the colors associated with the 4 distinct XANES clusters to the locations on the original sample that each spectrum came from, to produce Fig.7. This plot shows that the green (hematite-like) and red (hercynite-like) structures lie in contiguous geometric regions, and the blue cluster lies in a well-separated corner. This geometrical congruence alone serves as a sanity check on the DQC results, because geometrical proximity was not used as an input to the DQC analysis. The inset at the right shows a close up view of the blue cluster, making it possible to clearly see where we find the 60 metallic Fe plus magnetite pixels.

### 3.6   What Did We Learn?

Besides learning that DQC is capable of analyzing such large and complex datasets without using expert information or making assumptions about the type or number of clusters that may exist, we also show - for the first time - that datasets of this size and complexity can contain meaningful extended structures that are not isolated fixed points of the quantum evolution. Moreover, we showed that when such structures exist, it is possible to understand the kind of information they encode. This example also demonstrates that DQC can find small regions of interest within such big data sets because it is sensitive to small variations in the density of the data. Our results are in general agreement with a supervised analysis of the same data carried out by [6] looking for the best fit to a pixel spectral superposition of hercynite, hematite and pure iron. The information about the varying density of iron oxides and the matrix of other material is reflected in the assignment of different shades of red or green to each part of an extended structure. The fact that we were able to identify clusters of metallic iron and establish the unexpected presence of magnetite clusters, came as a pleasant surprise and demonstrates DQC's ability to reveal unexpected features of the data.

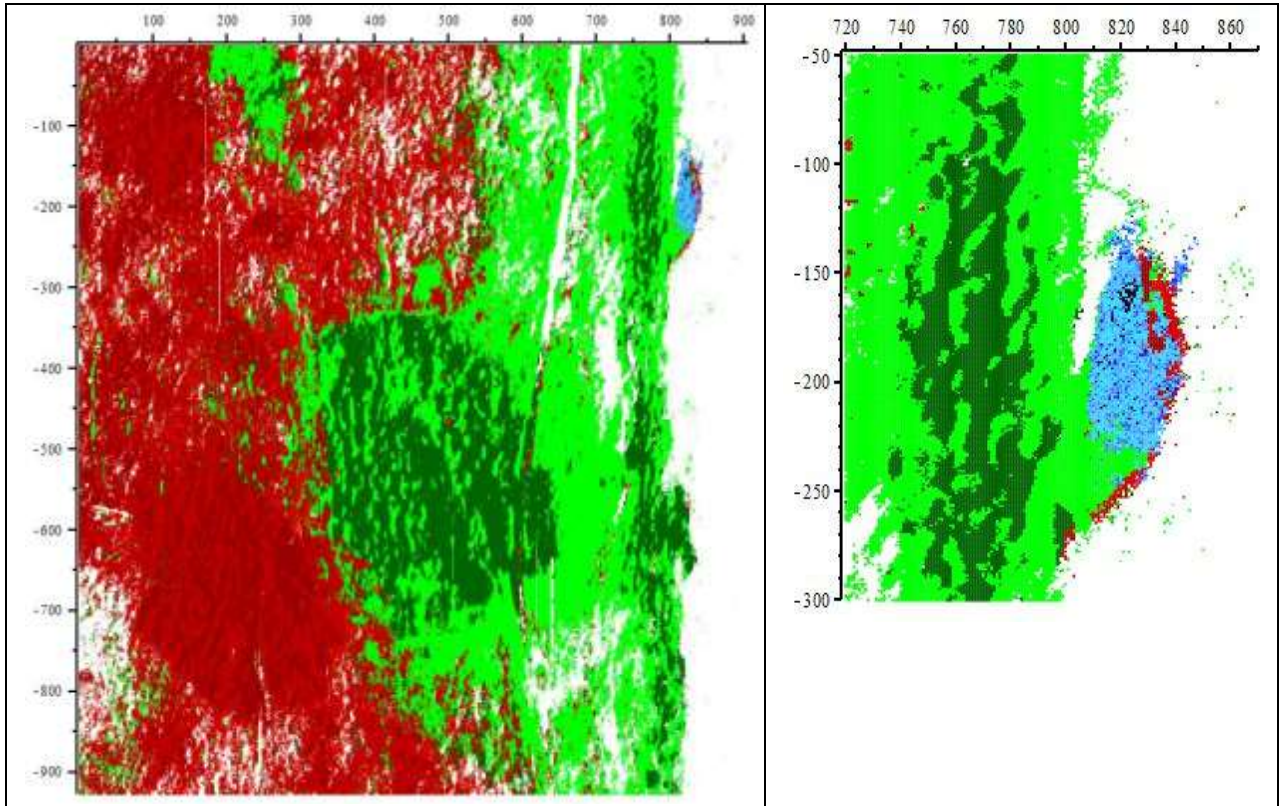

**Fig. 7**. A plot of the locations of the chemical phases according to the colors assigned to the structures seen in the evolution of the un-normalized raw data. The blue cluster is well localized and is shown in detail in the inset, using the additional colors employed in Fig.5. The dark blot on the inset on the right is the location of most of the 60 iron points, although a few single black dots are scattered throughout the region.

## 4. LCLS Pump-probe Data

### 4.1 Background

This dataset – that comes from the pump-probe facility at the SLAC Linear Coherent Light Source (LCLS) - provides another example of how DQC can be used to drill down and extract the content of a large noisy dataset. In this case it turns out that the challenge is to identify the 2.37% of the data that contains a useful signal. The DQC analysis demonstrates how to deal with data when DQC evolution results in an unusually large number of clusters that vary widely in size. As a bonus, we will show that in the final analysis both the good and the noisy data end up being useful. The good data sheds light on the physics of the sample, the noisy data ends up mapping the locations of possible defects in the detector.

In this experiment a germanium sample is *pumped* - by hitting it with an infrared laser pulse - and then *probed* - by hitting it with an x-ray pulse from the LCLS [7]. The purpose of the infra-red laser is to set the crystal into oscillation; the purpose of the x-ray beam is to take a snapshot of the crystal at a specified time after the infra-red laser pulse has been applied. Repeating this process a large number of times with different time delays between the laser and x-ray pulse produces a movie of the pattern of crystal oscillations. The aim of the experiment is to show that the resulting movie can be used to gain information about the quality of the data, and select the pixels that carry the information about the phonon spectrum of the sample. Since the beam parameters vary during the experiment most of the data is very noisy. Hence it is first necessary to isolate the small amount of good signal and then to show that the existence of sets of correlated pixels - with the same time dependence – implies that one is seeing coherent effects due to the laser pulse.

### 4.2 The Raw Data

The data is presented as a 262144 x 144 dimensional matrix. Each row records the time dependence of a single pixel in a 512 x 512 image of the germanium crystal. The first 40 entries in a row represent the time dependence of the scattered x-ray beam before the application of the infra-red pulse. Thus, each column of the data matrix is a picture of the scattered beam at a specific point in time.

To process the data we begin by removing rows where all of the entries are zero. This leaves a 204945 x 144 data matrix, which we rewrite – using an SVD decomposition - as $M_{data} = U\,S\,V^{tr}$. Figs.8A,8B and 8C display the rows 1, 2, 3 and 6 of $V^{tr}$. Note, the rows of $V^{tr}$ define 144 orthonormal time-dependent curves that – when added together - reconstruct the full data matrix. It is evident from Figs.7A and 7D that – as expected – there is very little signal between *t=1* and *t = 40*, since these times are before the infra-red pulse is applied. Fig.7B shows no strong difference between the region *t < 40* and *t > 40*, and Fig.7C shows no discernible difference at all. In fact, Fig.8C correlates well with known variations in beam parameters. While the average

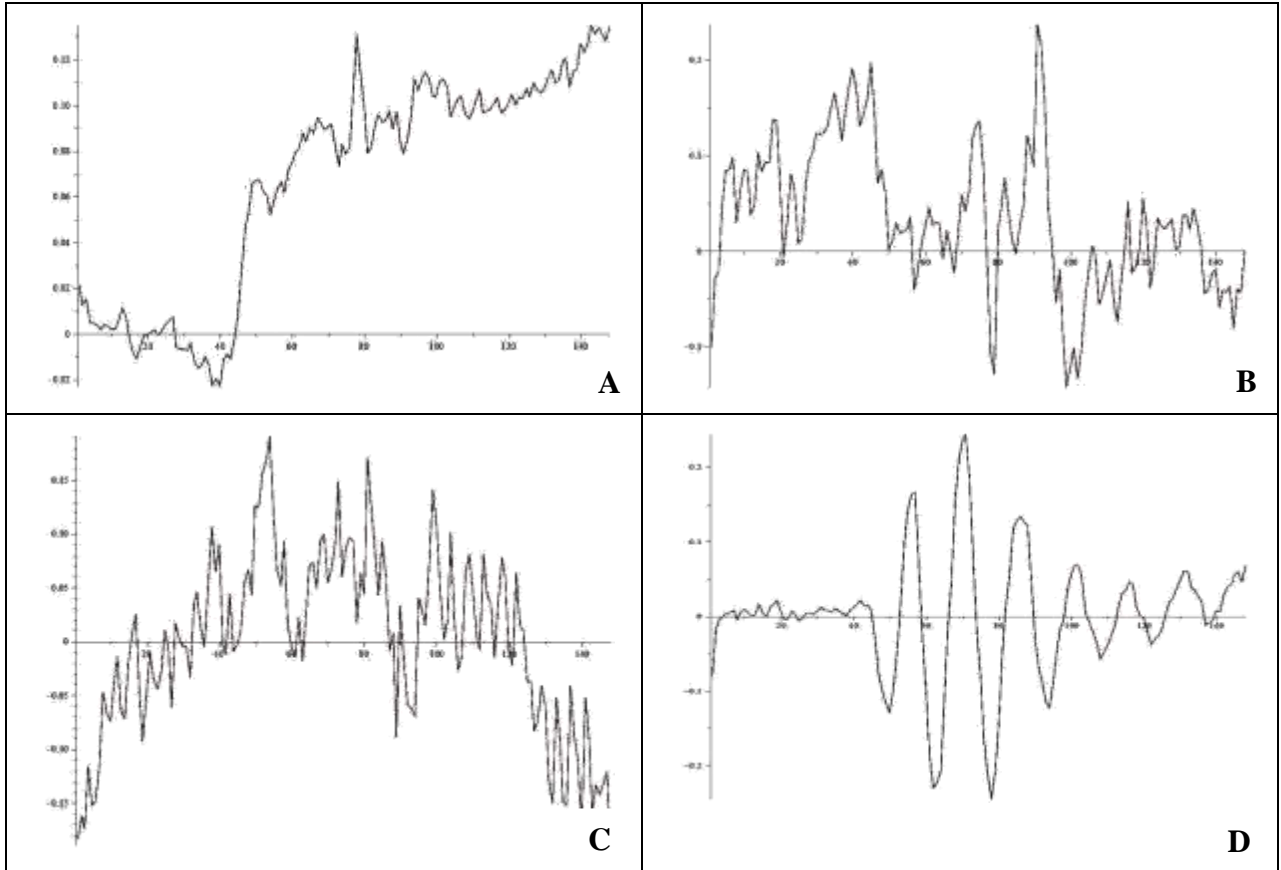

**Fig.8A-D** Plots of the SVD eigenvectors in rows 1,2,3,6 of $V^{tr}$.

signal – shown in Fig. 8A – shows no simple periodic behavior, we see that Fig.8D shows the kind of behavior one is searching for. To better understand why the average shows no sign of this behavior we turn to the plots shown in Fig.8A-D. By focusing on a single column of $U$ - which tells us how much the corresponding row of $V^{tr}$ appears in the data-matrix – we can construct an image of the crystal as seen by each of the SVD eigenvectors. Fig.8 shows four such images for the SVD components 1, 2, 3 and 6, corresponding to the time dependent curves shown in Figs.8A-D. Fig.9A exhibits two strong peaks (this signal comes from diffuse scattering from the tails of the Bragg peaks) and then a patterned small amplitude signal. The two strong peaks are what we expect. If the crystal is not excited by the laser then the x-ray beam should scatter from the crystal at Bragg angles (or – in reality – almost Bragg angles). Due to the excitation of phonons by the infra-red laser these peaks are broadened to produce the structures shown in Fig.9A. By the way, Fig.9B is very similar to what the average signal looks like if one does the same plot using only data for $t < 40$; *i.e.,* before the sample is pulsed by the infra-red laser. Thus, the signal that is seen in Fig.9A is what appears because the beam is scattering from the excited phonons. Figs.9B and 9C are constructed from the second and third SVD components whose time dependent signals (Fig.8B-C) have similar character both before and after $t<40$ – which is when the infra-red laser is turned on. They don't exhibit the kind of simple oscillatory behavior shown in Fig.8D that only exists for $t> 40$. Moreover, Figs. 9B-C don't exhibit strong signals related to the broadened Bragg peaks. This is particularly true for the third SVD component that is correlated with fluctuations in beam parameters and not with the applied infra-red pulse. Finally Fig.9D shows what happens for an SVD component that exhibits a definite oscillatory pattern. This strong signal is concentrated in the main peak and the smaller recurrent peak. Because Fig.8D shows significant time variation we expect the low-frequency modes to be absent and, thus, one will see no significant amplitude in the center of the peak in Fig. 9D. Indeed it shows four peaks that surround the central structure.

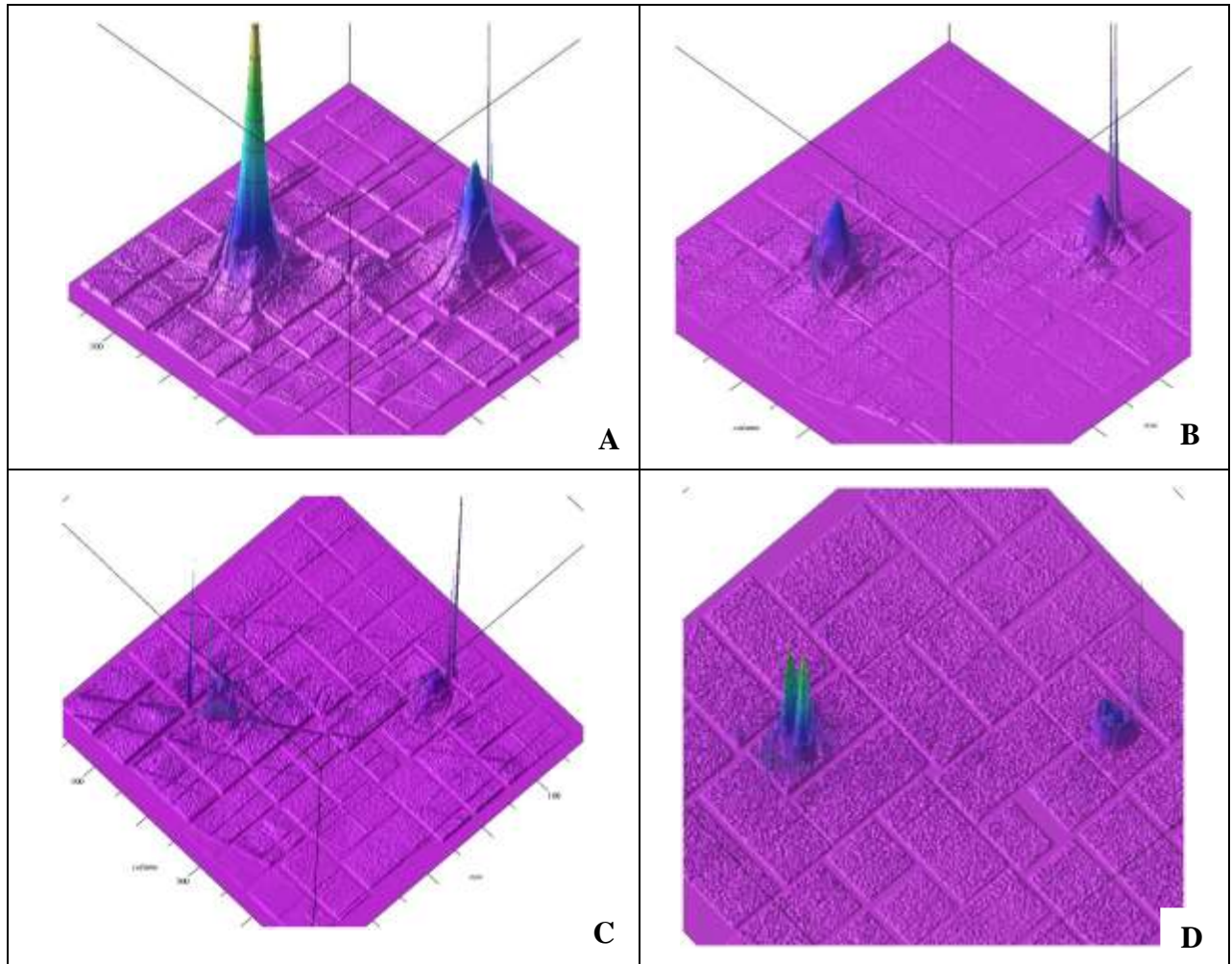

**Fig. 9A-D** Plots of images associated with SVD eigenvectors 1, 2, 3 and 6.

The patterned small amplitude signal outside of the peaks in all four figures is produced mainly by electronic noise in the detector and as such provides an image of the detector. Since over 95% of the data is this detector noise it is reasonable that the average of the data will not show the periodic variation that one should see from data concentrated in the peaks. Actually we will show that, due to the poor quality of much of the data in the peak regions, a much smaller fraction of the data will show a clear difference between data taken at times before the infra-red pulse is applied and times following the infra-red pulse.

While plotting the first few SVD components strongly suggests that we should focus on the data points that lie in the Bragg peaks, we nevertheless carried out a DQC evolution for all of the data in order to see if it contained any surprises. Since the results of this study confirm that the interesting data all lies in the two peaks this discussion has been moved to Appendix A. The data corresponding to the two broadened Bragg peaks amounts to ~15,000 data points which we re-analyze in the next section.

**4.4 The Second DQC Evolution**

The DQC analysis of the new smaller dataset begins with a new SVD decomposition of the data. Examination of the first SVD component, the component that represents the average of the time dependent data, reveals that it still looks like the average of the background data. Moreover, the second and third SVD components also seem to coincide with the second and third SVD eigenvectors for the full dataset. Since - from the point of view of the experimentalists - the real interest is in the variation of the data about the average behavior, we construct a *filtered* representation of the data. This is done by eliminating the contributions of the first three SVD components.

Initially - since there are only ~15,000 points in this filtered data - DQC evolution of the filtered data was performed in both 21 and 9 dimensions. These dimensions were initially chosen arbitrarily, however except for small variations, the results of both analyses agreed. Since working in 9-SVD dimensions takes less computer time, we only ran the 9-dimensional analysis for hundreds of frames. The final result is a very large number of small clusters and a small number of larger clusters. Since we have already seen plots showing the evolution of a complex dataset we relegated the plots showing typical frames in the evolution of this dataset to Appendix B. This Appendix also contains a brief discussion of novel features not seen in the analysis of the TXM-XANES data.

DQC evolution of the filtered pump-probe data results in a total of 669 point-like clusters that vary in size from clusters containing a single data point (these comprise slightly less that 10% of the data), to clusters containing tens of points and even a few clusters containing 500 - 1552 data-points. Since the infra-red pulse is always applied at $t = 40$ the signal seen in any spectrum for $t < 40$ corresponds to noise. For this reason we divide data into two types: *good spectra* that

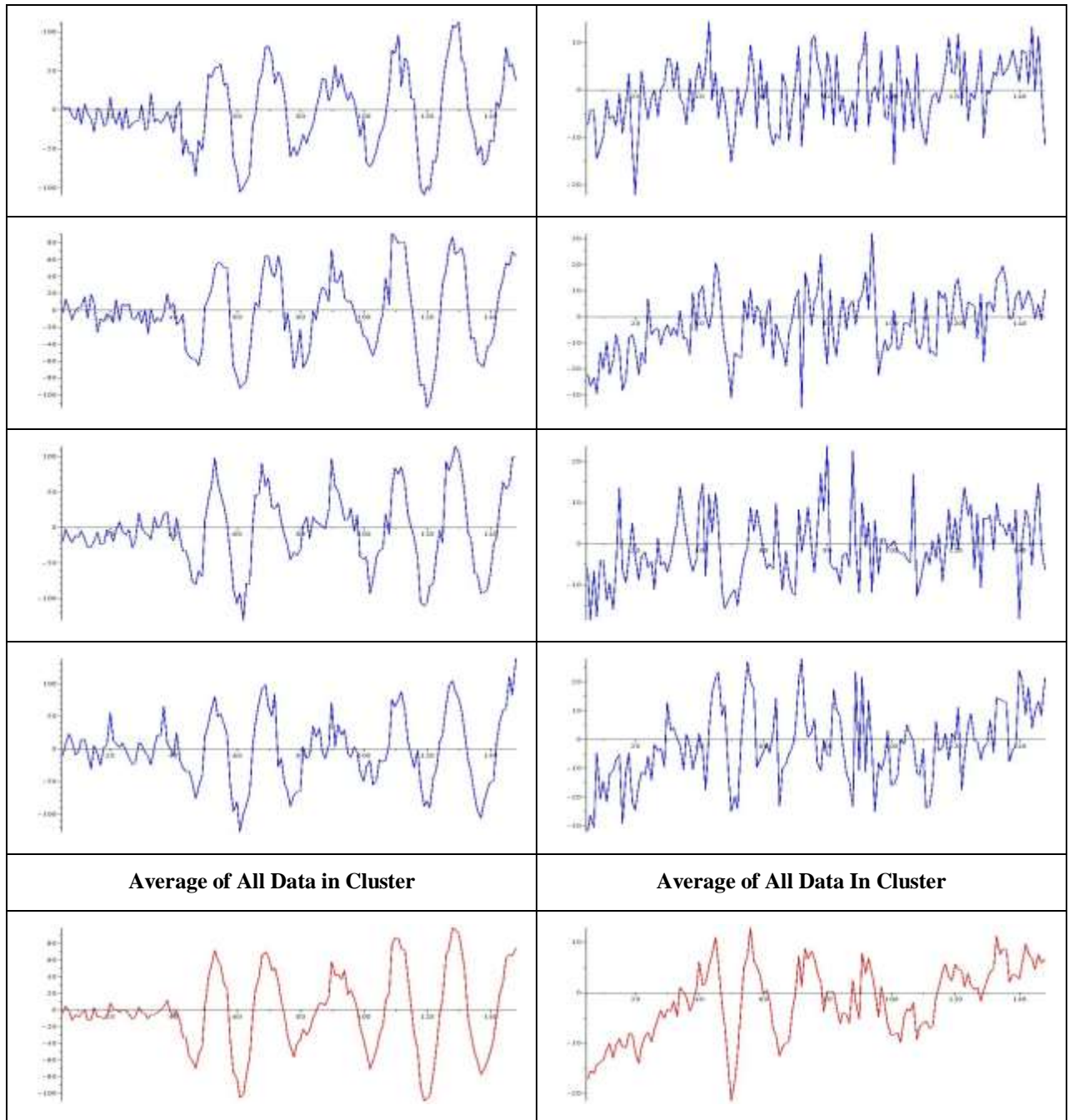

**Fig.10** The left column contains spectra for some points in a 9-point cluster. These show a strong signal, hence labeled as good data. The right hand column shows spectra which are identified as noisy data. Note that even for small clusters averaging produces a smoother curve.

exhibit strong signals for times *t > 40* and a smaller noisy signal for *t < 40* and *noisy spectra* where there is no appreciable difference between the amplitude of the signal for *t <* 40 and *t >* 40. Fig.10 shows typical examples of data taken from these two markedly different kinds of clusters. The left hand column shows examples taken from a cluster made up of good spectra and the right hand column shows examples from a cluster made up of noisy data, where the *t <*

*40* and *t > 40* signal have similar amplitudes. Each spectrum is shown two ways: the blue curves and red curves show the raw filtered data and the black curves correspond to the 9-D reconstruction of the same data. The close agreement between the raw data and the SVD reconstruction of the data explains why DQC evolution in both 21 and 9 dimensions give the same results. The fifth row in each column shows the average of all of the data in these clusters. It is evident from the plots that these averages are very similar to the individual examples. This demonstrates that the individual spectra in each cluster exhibit very similar behavior.

## 4.5 Separating Clusters into Good and Noisy Data

Investigating single spectra suggests a simple way to separate good data clusters from bad data clusters: namely, average the spectra in each of the 669 clusters and then – for each average – require that the mean value of the signal for *t <40* to be less than half of the average value for *t ≥ 40*. This procedure separates the 669 clusters into 367 clusters containing good data and 302 clusters containing noisy data. This leads to a total of 6172 spectra that correspond to good data, which is to be compared to the 12145 spectra associated with locations in the two peaks. Thus, we see that in the end only 2.37% of the original dataset contains good data.

This separation into good and noisy data will allow us to show that the oscillations seen in the good data exhibit correlations between phonons in disjoint regions of the broadened Bragg peaks. This kind of correlation implies that good data are capturing the coherent excitation of the germanium lattice by the infra-red laser and thus, they can be used to study the phonon spectrum. Before demonstrating this result it is worth demonstrating that the noisy data also has something to tell us about the detector and the quality of the cut we made to isolate the diffuse Bragg peaks.

By comparing plots of the good and noisy data when both are plotted on the two-dimensional plane of the pixel-detector we made a serendipitous discovery. The left hand plot of Fig.11 has a dot at the location of every pixel associated with a good signal. The right hand plot does the same but for pixels that contain noisy data. As is evident from these two plots the good and the noisy data are intermixed to a certain degree. In other words, the noisy data come from detector pixels that are adjacent to and between pixels that contain good data. Presumably this means that not all of the pixels on the detector are working well. This assumption is buttressed by the fact that the left hand plot shows a small circular region - at the bottom of the region corresponding to the main broadened Bragg peak - that contains no good data. Moreover, the right hand plot has a solid circle of noisy data in the same region. This leads us to conclude that this circular region represents localized damage to the detector. If we further examine the two plots we find that the strongest signal that appears in the good data corresponds to locations where there is an absence of signal in the noisy data. Regions where the x-ray scattering signal is far above the detector noise should produce pixels with the cleanest signals and it does. In the peripheral region, where the x-ray scattering signal drops towards the amplitude of the detector noise, we expect to find the noisier pixels. The separation of good data from noisy data by extracting the clusters produced by DQC evolution and averaging the signals in each cluster to reduce stochastic noise, allows us to create a simple criterion for separating good from noisy data that does a much better job of preserving the useful signal.

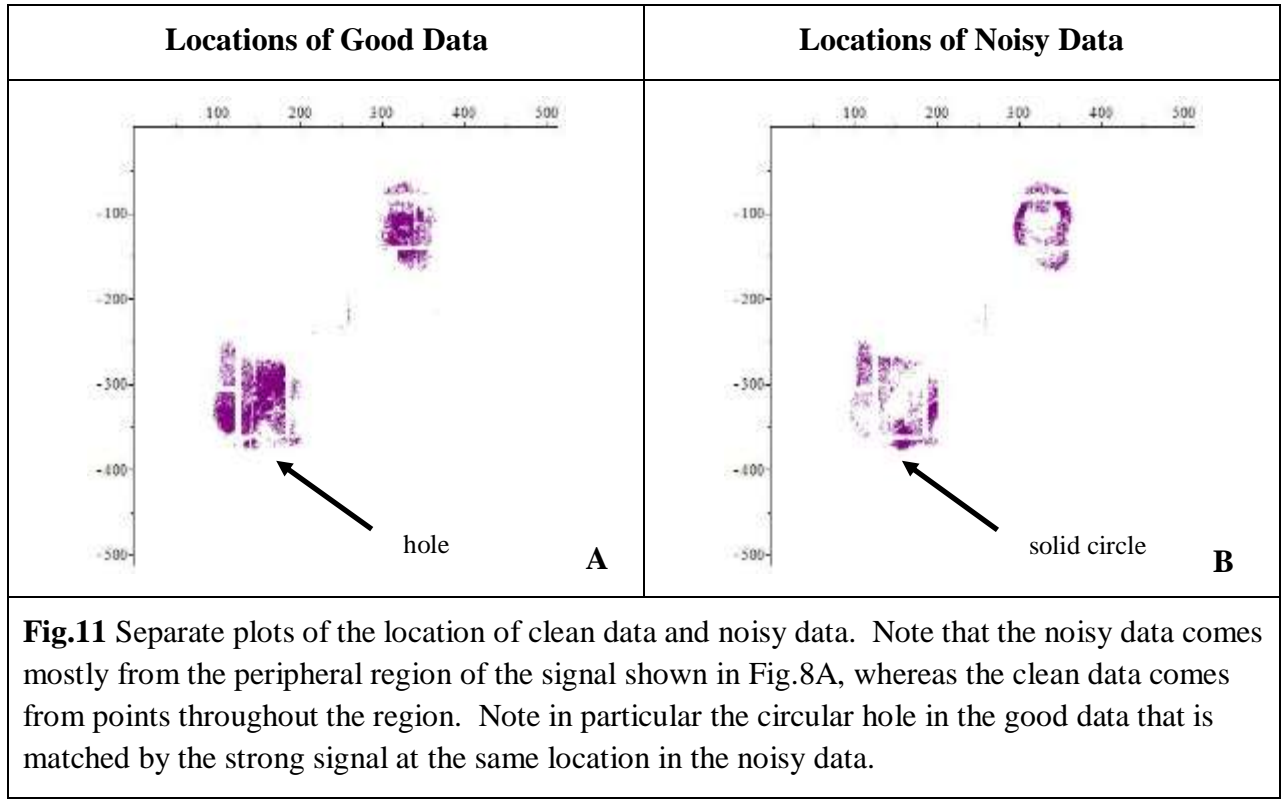

**Fig.11** Separate plots of the location of clean data and noisy data. Note that the noisy data comes mostly from the peripheral region of the signal shown in Fig.8A, whereas the clean data comes from points throughout the region. Note in particular the circular hole in the good data that is matched by the strong signal at the same location in the noisy data.

### 4.6 The Evidence for Correlations

Fig.12 shows a plot of points in three of the good clusters. Points corresponding to the same cluster are shown in blue, cyan and magenta. As is evident from the plot – especially in the main peak – the blue and cyan points are distributed almost symmetrically about the center of the peak. This is, of course, what one would expect for coherent excitation of the lattice, since phonons with opposite spatial momenta should show the same time dependence. Since displacement from the center of the peak is correlated with phonon momentum exhibiting separated pixels with the same time dependence is evidence for the fact that these phonons are part of a coherent signal. The fact that spectra exhibiting the same time dependence are not precisely opposite to one another on the plot has to do with the preparation of the sample. The only reason we are able to exhibit these correlations so cleanly is that we are able to cluster signals from individual pixels according to their actual time dependence rather than trying to work with Fourier components.

### 4.7 What Have We Learned?

First, the main scientific result is that data obtained using this pump-probe technique can be used to study coherent lattice vibrations and thus the non-equilibrium phonon structure of the crystal. This was accomplished despite the noisy nature of the data, problems with the detector and problems introduced by the difficulty in stabilizing the properties of the x-ray beam. Since the main goal of this exercise was to prove that this kind of analysis can be done, we conclude that the DQC analysis was very successful.  Furthermore, the fact that DQC allows us to establish correlations between pixels by looking at data in the time domain – rather than in the Fourier domain – opens up the possibility of recovering more detailed information about the non-equlibrium dynamics in the crystal.  This result means that it should be possible to carry out this sort of study on crystals whose phonon band structure is not well understood.  We also found that by separating good and noisy signal using DQC we could identify regions where the pixel detector failed to work properly.

From the data mining perspective we see that once again DQC successfully clustered complex, noisy signals by their structure in feature space.  In this case we extracted from data coming from pulse probe experiments the 2.37% of the signals that contain useful information.  A new feature of this analysis is that the final number of clusters - 669 in all - is quite large and the final clusters vary enormously in size.  This is not what one might have expected going in to the analysis.  Since there is no way to guess how many significantly different spectra would be visible in the data, it was important that DQC – unlike many other methods - doesn't need to make any a-priori assumptions about the number of clusters that may exist.  Another important observation is that by performing the initial analysis in both 9 and 21 dimensions we found that the two analyses agree with each other, thus we concluded that it was safe to work in the lower number of dimensions.  This is easy to do because the time required for a DQC analysis only grows linearly with the increase in dimension.

**Some Good Data Colored According to Cluster**

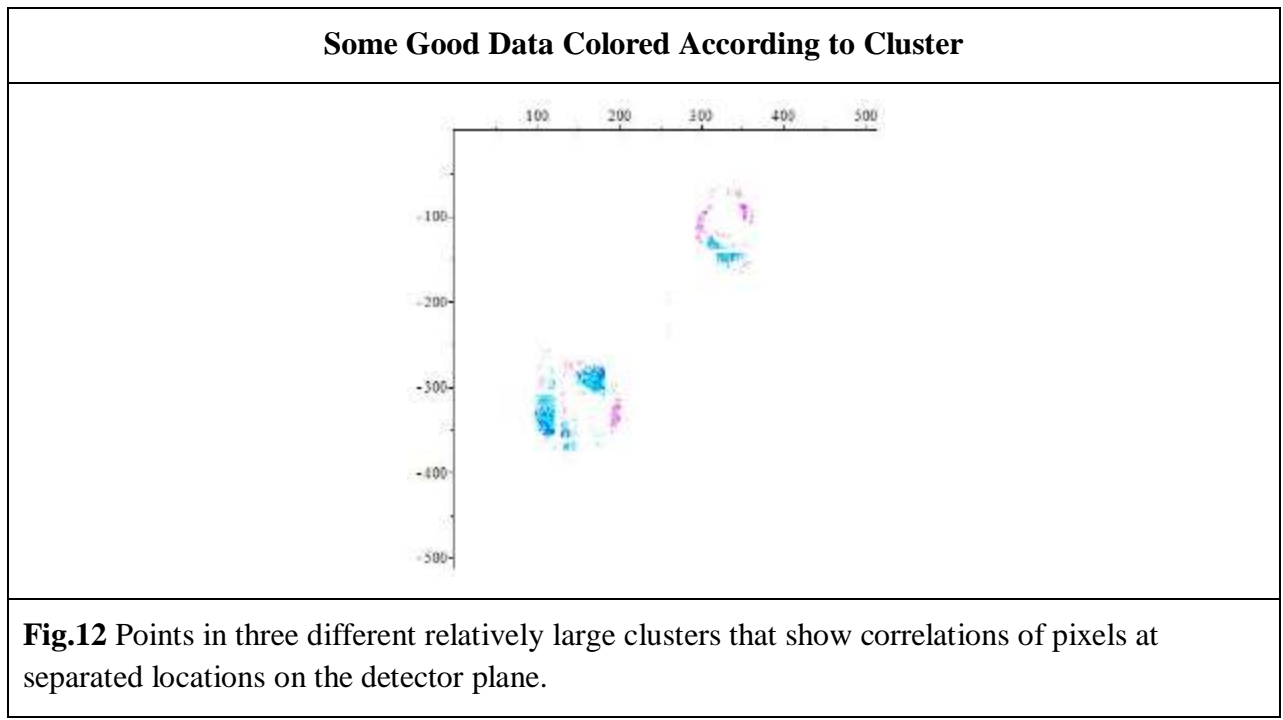

**Fig.12** Points in three different relatively large clusters that show correlations of pixels at separated locations on the detector plane.

## 5. Aquaporin Data

The third dataset presents an entirely different kind of challenge and demonstrates a different mode for applying the DQC algorithm. Cells exchange water with their environment faster than can be accounted for by diffusion of water molecules through the cell wall. The explanation of this phenomenon is that the cell membrane contains pores that allow water and other molecules to move in and out of the cell body. These channels are created by proteins called aquaporins. Genetic defects involving aquaporins have been associated with serious human diseases. Understanding how and why aquaporins selectively transmit either water or glycerol molecules could thus lead to technologies for alleviating these medical conditions.

### 5.1 The problem

Crystallizing a protein, in order to obtain its 3-dimensional structure, is the usual first step in figuring out which locations along the chain of amino-acids making up the protein determine its biological function. Unfortunately, most proteins are hard or impossible to crystallize. It would be a huge advance if identifying important locations along the protein's amino-acid chain could be done without using 3-D information. . Conventionally one looks for such clues using a Multi-Sequence Alignment (MSA) matrix, assembled by using similar proteins performing the same functions in different species. We show that given such data DQC can be used to find which locations on the protein are responsible for its classification into a water or glycerol transmitter.

### 5.2 The Data

The data consists of 529 aligned amino-acid sequences for two different kinds of aquaporins. [10]. The functional difference between these proteins is that one creates a channel that passes water molecules, and the other creates a channel that passes glycerol molecules. Each row in the data-matrix specifies the amino-acid sequence for a particular protein. Hence, since there are 20 amino acids, each row of the data matrix is given as a string of letters: *i.e.,* A, C, D, E, F, G, H, I, K, L, M, N, P, Q, R, S, T, V, W, Y. We convert these letters to numbers from 1 to 20, so a single row contains 192 integers. Since some locations correspond to gaps in the aligned protein sequences and we arbitrarily assign the non-integer value 10.5 to those locations. The choice of this arbitrary number is unimportant and doesn't affect the following analysis.

### 5.3 The DQC Analysis

The analysis begins with an SVD decomposition of the data matrix and its dimensional reduction to the first three principal components. The first step in the analysis is to determine whether the data contains the information needed to distinguish a water aquaporin from a glycerol aquaporin. Fig.13A shows the original data ($U$-matrix) plotted in SVD dimensions 2-3 where the extended structure is most apparent. Points are colored according to the identification of the protein as either a water or glycerol aquaporin. This plot is quite diffuse, but the separation of the red and green points suggests that separating the two types of aquaporins should be possible. Fig.13B shows the result of applying DQC evolution to the data. The plot shows an extended *v*-like structure containing three distinct parts where two contain mostly red or mostly green points. This structure shows that the information needed to distinguish the proteins is present in the data. However, the extended structure suggests that the data is very noisy. The challenge is to refine this separation and identify which of the 192 locations ($V^{tr}$-matrix data) are most important to this classification.

Since the extended structure is most apparent in SVD dimensions 2 and 3 we turn our attention to the SVD $V^{tr}$-eigenvectors corresponding to these two values. Since the entries along the rows of $V^{tr}$ correspond to the different locations along the protein's sequence, the idea is that we can find the locations that are most relevant to the classification, by selecting from rows 2 and 3 of $V^{tr}$ the locations that contain the largest numerical values. By plotting the numerical values appearing in row 2 and row 3 of $V^{tr}$ we find that in each row the numbers naturally separate into three parts. Selecting those locations containing the largest values we obtain a list of 30 locations to be used to continue with the analysis. Using DQC it is a simple matter to check that this approach makes sense. Fig.14A and 14B show the result of removing the unimportant features (162 locations along the sequences) from the dataset and re-plotting the data. Before DQC evolution the restricted data is still diffuse: while it appears to exhibit better separation between the glycerol and water aquaporins (red and green points) it is hard to be sure that removing the so-called unimportant features made a significant difference. However the story is very different after DQC evolution. After evolution the data is seen to cleanly divide into two compact clusters, each of which is almost entirely red or green. The small number of apparently incorrectly classified proteins is consistent with the known probability of errors in the data. These plots show that the features deemed unimportant - and thus removed from the data - play little or no role in distinguishing "water" and "glycerol" aquaporins. Given such a small number of interesting features, it is a simple matter to examine histograms showing how many times a given amino-acid appears at a specific location. These histograms identify four locations at which a specific amino-acid appears more than 70-80% of the time. Figs.15A and 15B exhibit histograms for location 161, while Figs.15C and 15D give the same information for location 173. The four locations identified in this way seem to be most effective at uniquely identifying the two kinds of aquaporin proteins and so they are the best candidates for locations that should be manipulated to produce a therapeutic result. A correlated sequence analysis (CSA) of this data [11], with a careful choice of parameters, identified the same locations as important. However, the DQC analysis was needed to set the direction. Moreover, it is much simpler, much faster and quite convincing even before a rigorous statistical analysis is applied.

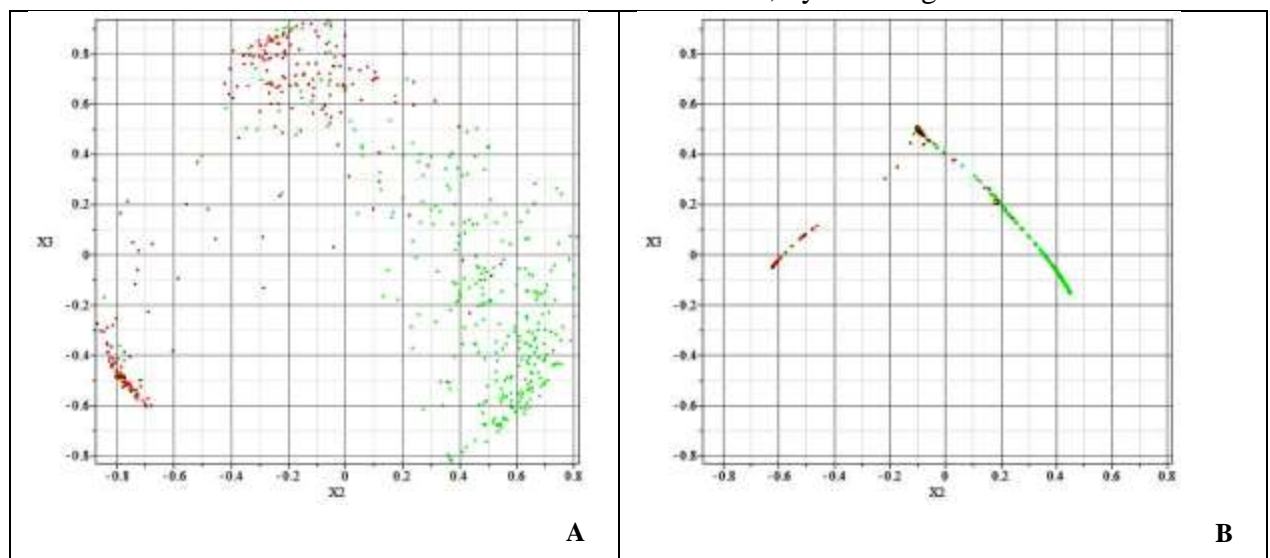

**Fig.13** (A) The original, diffuse data. (B) The same data after DQC evolution.

## 5.4     What has been learned from this example?

This example shows how DQC can be used to attack a problem where, from the outset, one has a classification of items in the dataset, but one doesn't know if the measured data contains the information needed to explain this classification. This kind of problem is common to many fields. As another example, assuming that one has the full genome and medical history for an individual, can one predict from the genome data whether the person has, or will have, a particular disease? This analysis shows that once one knows that the necessary information is present in the data, one can identify unimportant features by the fact that eliminating them improves the clustering results. Once one reduces the problem to a manageable set of features, there are many ways to home in on the answer.

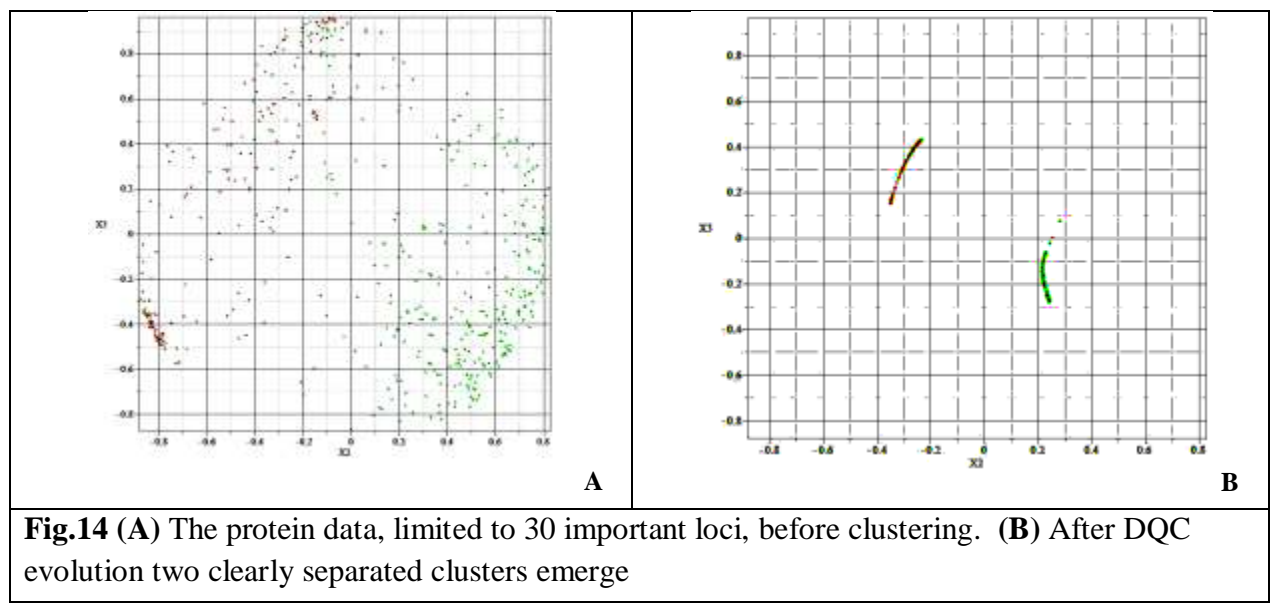

**Fig.14 (A)** The protein data, limited to 30 important loci, before clustering. **(B)** After DQC evolution two clearly separated clusters emerge

## 6.     Earthquakes in the Middle East

## 6.1     The Problem

Earthquakes are conventionally labeled by magnitude, location and time of occurrence, although other physical characteristics can be included. The question is if it is possible that earthquakes show clustering if one considers only physical parameters and ignores location and time of occurrence information, and moreover can it lead to novel insights?

## 6.2 The data

The data comes from a catalog [12] of 5693 earthquakes that occurred in the Eastern Mediterranean Region (EMR) and the Dead Sea Fault (DSF) over a 20 year period. In addition to location and time of occurrence, this catalog contains five physical parameters for each earthquake: magnitude of the earthquake, seismic moment, stress drop of the fault due to the earthquake, source radius and corner frequency (beyond which the spectrum is white noise).

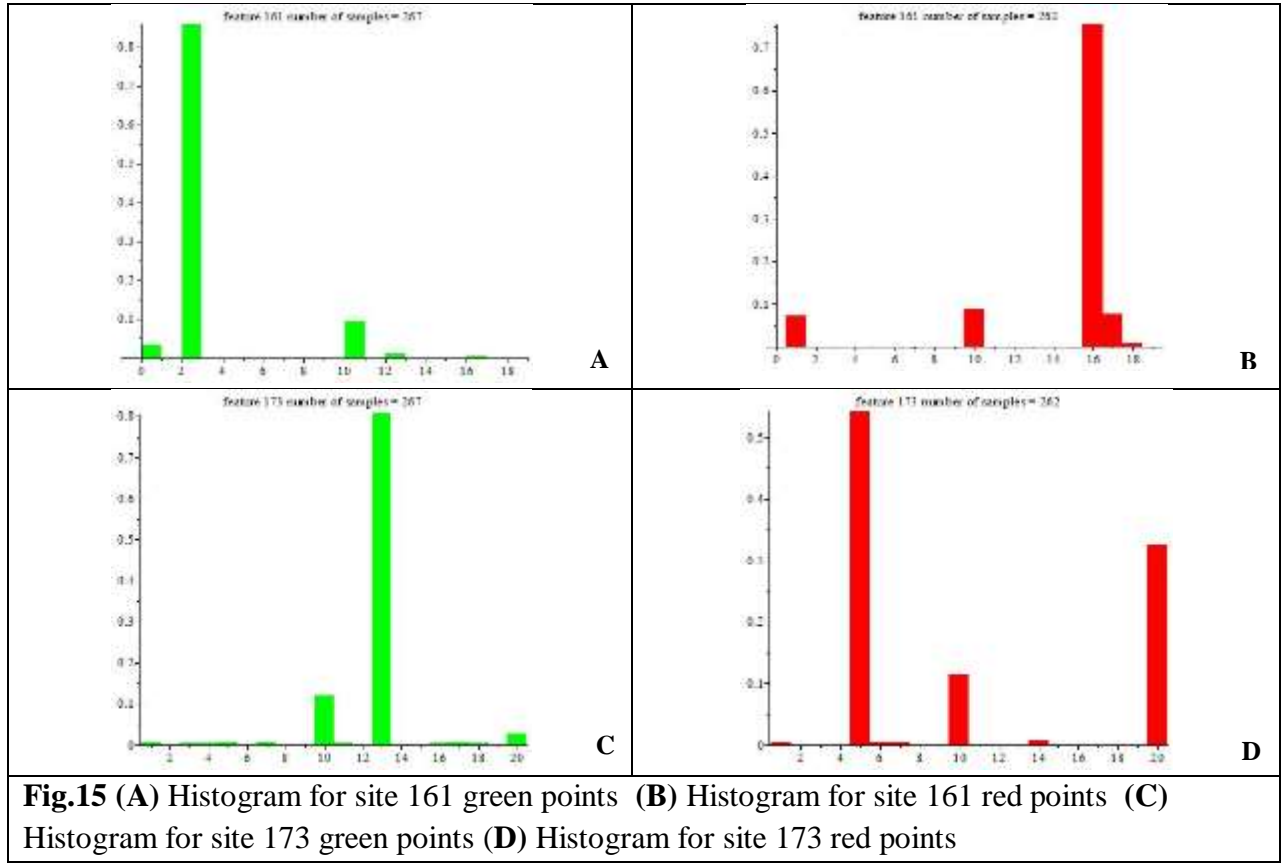

**Fig.15 (A)** Histogram for site 161 green points  **(B)** Histogram for site 161 red points  **(C)** Histogram for site 173 green points **(D)** Histogram for site 173 red points

## 6.3 The DQC analysis

DQC analysis was performed on a 5-dimensional SVD transformation of feature space. The data points eventually converged onto a few clusters. To exemplify the clustering process we present in Fig.16 the initial directions of motion of the data-points, in the form of unit vectors of -gradV. Note that no obvious separations that indicate cluster boundaries exist in the data. Nonetheless the movement of the data-points is dominated by centers of high density of events which become fixed points of the dynamics of our clustering procedure.

These clusters were obtained from information of the distribution of events in parameter space without any reference to the temporal and geographic location of the earthquakes. The combination of the DQC results with time and location information revealed interesting correlations with known faults. Fig.17 shows the location of the various earthquakes, colored according to cluster, on a map of the Middle East revealing the surprising result that almost all events of the orange cluster are concentrated geographically around the Gulf of Aqaba. An analysis of their distribution in the original parameter space shows that these earthquakes are of low magnitude but possess relatively large stress-drop values. The central values of the parameters of all clusters are displayed in Table 1. Moreover, it turns out that almost all orange events have mostly occurred within a few months following the strongest earthquake that took place on Nov. $22^{nd}$, 1995 in the Gulf of Aqaba. In hindsight one may now assign a geophysical meaning to events of the orange cluster: they represent major ruptures that have occurred following the strongest earthquake. Indeed the analysis of events in the relevant period carried out by Baer *et al.* (2008) [13] observed slip distributions which were quite unique to just this period.
For more details of the QC and DQC analyses of these data see [14].

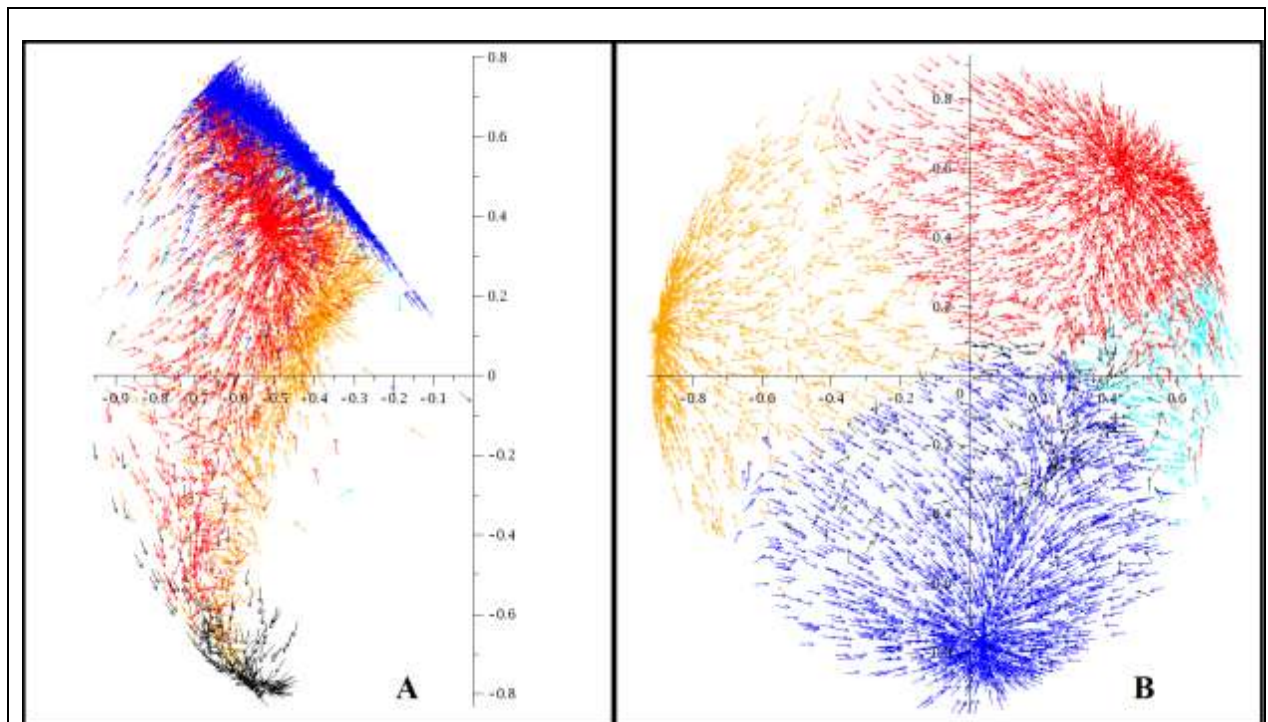

**Fig.16:** Unit vectors showing the negative gradient of the potential in SVD space. Places where all of the arrows point inwards reveal clear centers of attraction. **(A)** projection of gradient into dimensions 1-2. **(B)** projection of the gradient into dimensions 3-4. The colors of the arrows correspond to the colors assigned to the clusters obtained from the clustering procedure (see Table 1).

### 6.4 What have we learned from this analysis?

We see that even a relatively small dataset exhibits surprising, non-trivial density variations that are not evident in original data. The correlation of clusters obtained from the physical parameters with the time and location information was a surprise, leading to a unique interpretation of the orange cluster. The hope is that such analyses will be adopted as acceptable tools by the seismologic community, leading to further insights that can be derived from such data.

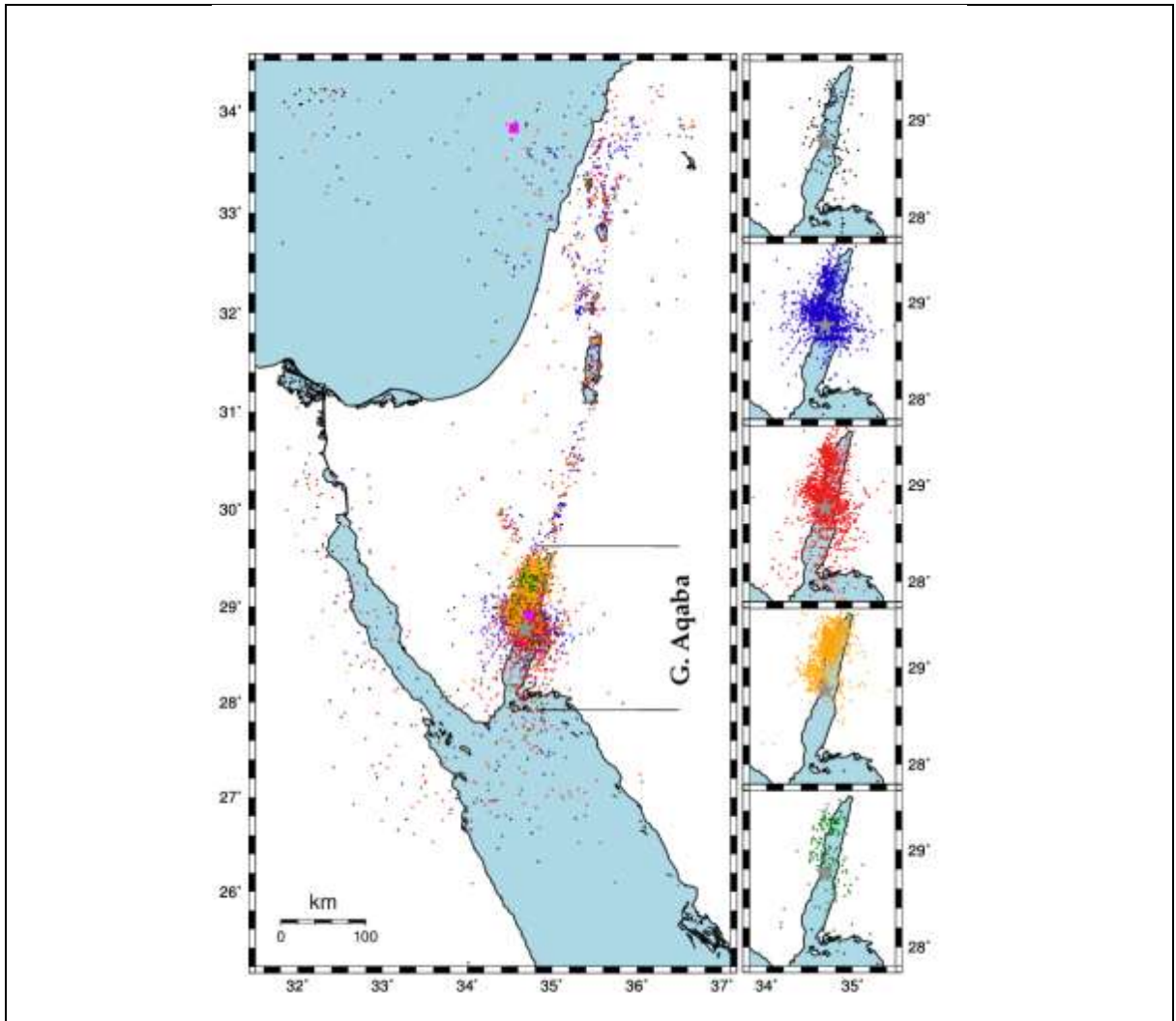

**Fig.17:** earthquake events, classified by the assigned clustering colors (see Table 1). The region of the Gulf of Aqaba is marked on the regional map on the left side. Detailed insets associated with the major clusters within the Gulf of Aqaba are shown on the right side. This figure was prepared using the GMT program [15]. The major earthquake of Nov. 22$^{nd}$ 1995 is denoted by a grey star on the figure and all insets. Two magenta events, corresponding to the cluster of 2, are depicted near the major earthquake of 1995 and within the Med. Sea.

| Clusters | Magnitude | Stress drop (bar) | Corner frequency (Hz) | Source radius (km) |
|---|---|---|---|---|
| 2052 blue | 2.0 | 2 | 4.05 | 0.32 |
| 2012 red | 3.4 | 9 | 3.33 | 0.41 |
| 1142 orange | 1.9 | 13 | 7.78 | 0.17 |
| 229 black | 3.8 | 62 | 3.46 | 0.44 |
| 255 cyan | 3.1 | 4 | 2.49 | 0.54 |
| 2 | 2.8 | 12 | 4.51 | 0.27 |
| 1* | 7.2 | 69 | 0.17 | 20.0 |

**Table 1:** Average parameter values of earthquakes within the different clusters. First column enumerates events in clusters. Colors are assigned to clusters containing many events. 1* is the major earthquake of Nov 22nd, 1995 (following [15, 16]).

## 7. Financial Data

The problem of studying stock market data to reveal interesting structure has a long history. In this section we look at data coming from the Standard and Poor's (S & P) 500 list and discover that it exhibits unexpected, interesting structure. The data we work with consists of daily stock prices recorded throughout the period January $1^{st}$, 2000 – February $24^{th}$, 2011, comprising a total of 2,803 active trading days. It is worth noting that this data includes the crises of 2002 and of 2008.

### 7.1 The Problems

In what follows we limit attention to the 440 stocks that exist throughout the recorded period. This data can be discussed in two different ways. First, since these stocks are conventionally classified as belonging to one of 9 *market sectors*: 1. Basic Materials, 2. Communications, 3.Consumer, Cyclical, 4. Consumer, Non-cyclical, 5. Energy, 6. Financial, 7. Industrial, 8. Technology, 9. Utilities, we can ask if clustering the stocks according to the time history of their daily returns will conform to this a priori classification. The answer to this question – unsurprisingly - is "Yes!" For further details see [13]. Second, we can turn the problem around and ask if there is a pattern of days for which the *market snapshot* of all stock daily prices reveals an interesting temporal structure. Surprisingly, the answer to this question is also "Yes!" Here we will deal with the second problem.

## 7.2 The DQC Analysis

The data for this analysis is a matrix, **P**, which has 2803 rows and 440 columns. Each row of the matrix corresponds to one day stock prices for all 440 S&P stocks for which we have data. All daily prices are normalized to the price of each individual stock on the first day of the data. As before, we perform an SVD-decomposition of the matrix **P** and then dimensionally reduce the problem to 10 dimensions (projecting the points onto a unit sphere). Next we cluster the data using DQC.

The results of the analysis, displayed in Fig.18, show the existence of clusters of days –indicated by the solid bars. We refer to these clusters as *market epochs*. Clearly there are only three epochs during the first five years (with only one covering the 2002 crisis), and then - for the second half of the studied period (including the 2008 crisis) – a larger number of shorter epochs.

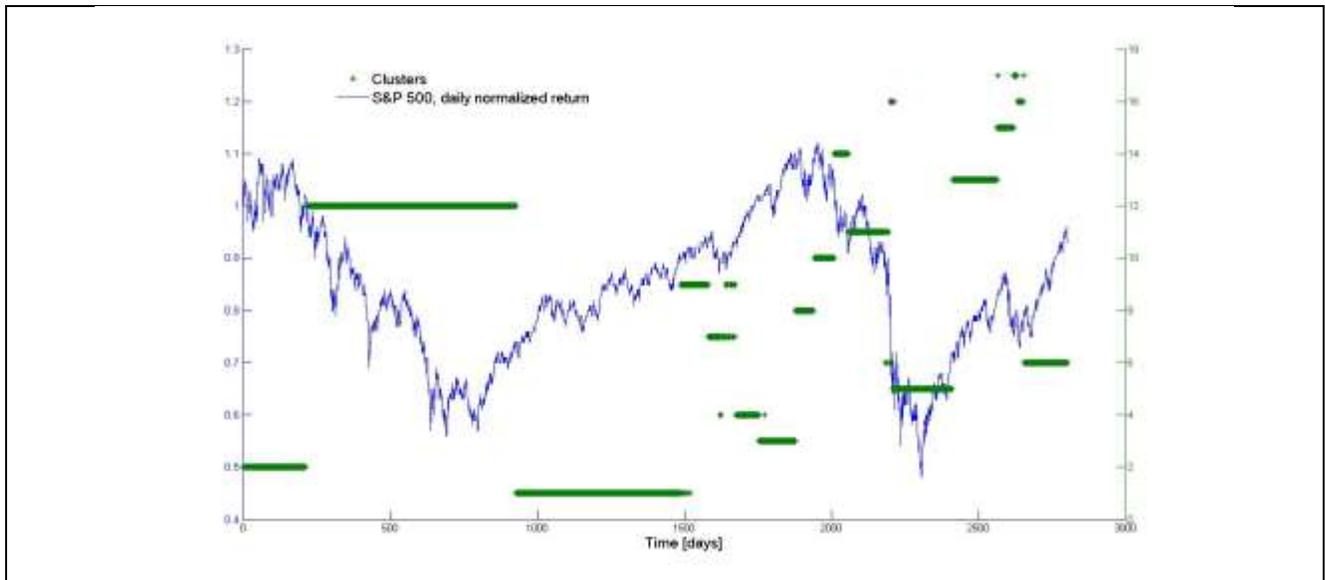

**Fig.18**. Temporal DQC clustering of the matrix **P** into 17 epochs, represented by bars. For comparison we plot the S&P 500 index for the same days, just to serve as an indicator of the known market behavior with its crises of 2002 and 2008.

We will now show that each temporal cluster has its own unique characteristics. One way to display the difference between the epochs is to display daily prices of stocks averaged over different sectors. This leads to results of the kind displayed in Fig.19. Each day is represented by a point on the 3-dimensional plot whose coordinates are given by the average price of a stock in each of the three dominant market sectors; *i.e.,* Basic Materials (1), Energy (5) and Industrial (7). By displaying each of the days that belong to one of the 17 epochs in different colors we see clear structure in the data. An alternative - perhaps more informative - way of exhibiting the differences between epochs is to study the Pearson correlations between the return values of stocks during each one of the 17 epochs. The result of computing these matrices is displayed in Fig.20. Note that in constructing these matrices stocks that belong to the same market sector lie near one another. Furthermore, the order of the individual matrix plots corresponds to the order of appearance of the epochs. From these plots we see that the first three epochs that cover more than half of the temporal span (from 2000 to 2005) show significantly lower average correlations among stocks of different sectors than during the following epochs. Moreover we note the overall phenomenon that the strength of correlation increases in time. The unique character of each of the 17 epochs is clearly demonstrated by the differences between the average correlation matrices.

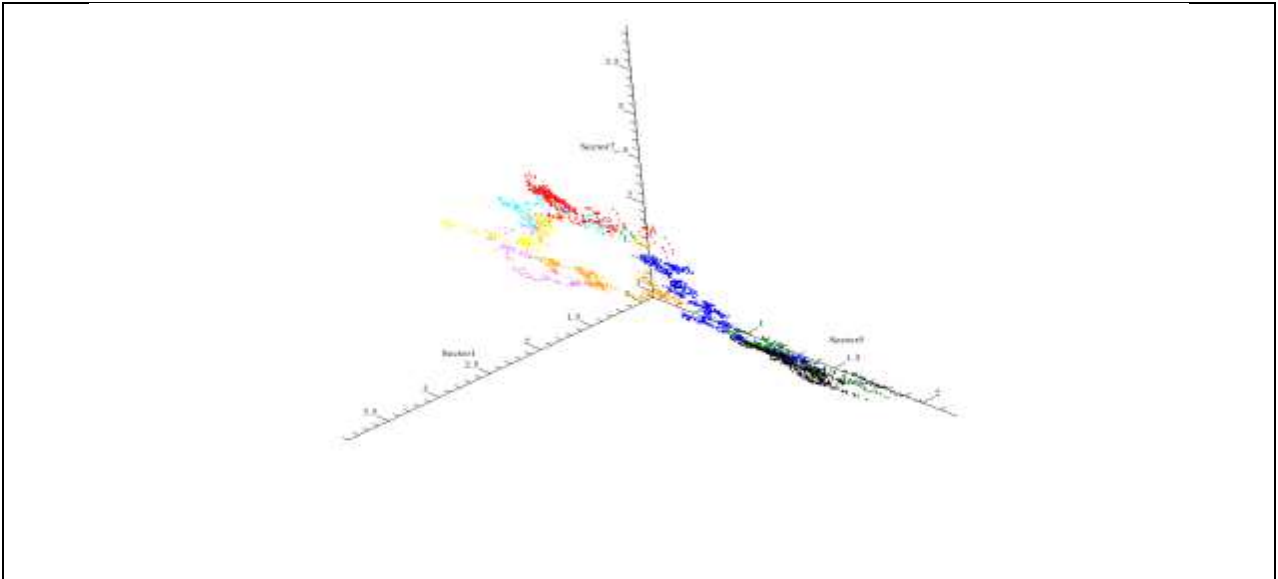

**Fig.19** Average daily prices for three sectors (1,5,7) are plotted in a space spanned by these sectors. The data are seen to cluster into different epochs, distinguished by the different colors.

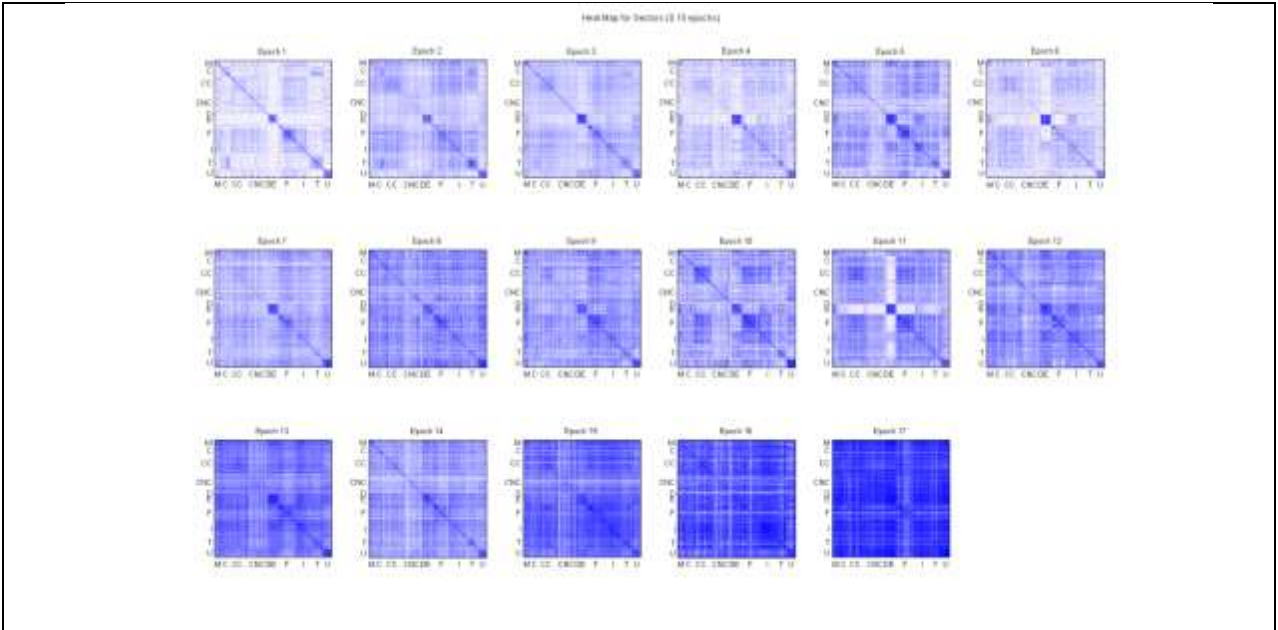

**Fig.20** Pearson correlations among all the different stocks for the each one of the 17 epochs. The sectors are abbreviated as M=Basic Materials, C= Communications, CC=Consumer, Cyclical, CNC= Consumer, Non-cyclical, E= Energy, F= Financial, I= Industrial, T= Technology, U=Utilities. The order of the displayed by heat-map matrices (darker implies stronger correlations) corresponds to the temporal ordering of the epochs.

### 7.3 What have we learned from this analysis?

While this data is not especially large, the analysis demonstrates how the same data can be looked at in two totally different ways. First, as already noted, clustering the stocks on the basis of the return matrix reveals the existence of well-known market sectors. Second, we showed that by flipping the rows and columns and using the daily stock prices (rather than the rate of return) we saw that the behavior of the total market (at least as represented by the 440 S&P stocks) exhibited temporal clustering into epochs. The dual relationship between information stored in the rows and columns of the data matrix presumably holds for dynamical systems described by time-series, in analogy to stock prices *vs* time. In this case the division into epochs revealed similarities between stocks in particular epochs, which could not be deduced from correlations evaluated over long time-scales.

It is worth noting that in a recent paper, Munnix et al [17] proposed using correlations between stocks, over short time periods (e.g. week or month), to define *states of the market*. In other words, different states will be associated with considerably different correlation matrices. Our proposal of clustering the temporal domain into epochs is an alternative approach: each epoch may be viewed as a state of the market realized within stock-space. These two methods may disagree about the specific division into epochs, because they are based on different mathematical manipulations of the data, but both serve as alternatives for discretizing the complex system into categories that allow for a quantified analysis of the market structure.

### 8. Other Applications

It should be apparent from the variety of examples discussed in the preceding sections that DQC applies to all sorts of data. In addition to the analyses presented here, DQC has been used to study stock-market data, single nucleotide polymorphism data (SNPs) for Alzheimer's disease, particle physics data and document classification. It remains to be shown if the same methodology can be used to index datasets that are comprised of pictures; *e.g.,* databases of diffraction patterns collected in the study of catalysts, or faces. These studies and possible future extensions demonstrate that DQC can be important to people working in such diverse fields as chemistry, biology, particle physics, astrophysics, genomics, proteomics, business, finance, analysis of social networks and national security.

### 9. Summary

DQC has demonstrated a powerful ability to expose hidden structures and determine their significance. This ability is - to our knowledge - unique to DQC. Combining this with the fact that DQC is both data-agnostic and completely unbiased, creates a powerful new way of visually interacting with poorly understood data. We demonstrated the power conferred by this ability in three examples of complex datasets - drawn from different fields of science - that contain hidden extended structures that encode important information. This fact that such structures often exist in large, complex data is an important new insight. In the case of the TXM-XANES data, the extended structures encoded information about the density of the iron-containing compounds at each point in the sample and the presence of those x-ray absorbing compounds that do not contain iron.  The full analysis proved that the structures were a feature of the data and not an artifact of DQC evolution.  In the case of the LCLS pump-probe data a single extended structure was shown to contain most of the background.  By identifying the origin of this structure we easily separated the over 95% of the data that contained no interesting signal from the important data.  Further study then extracted the 2.37% of data containing a good signal and showed that using DQC to analyze this kind of data it will be possible to study the phonon spectra in less well understood crystals.  In the aquaporin example, the appearance of an extended v-like structure in the first stage of clustering provided the clue that allowed us to do feature selection and reduce this difficult problem to a size that could be managed by standard means.  In contrast, the earthquake example demonstrates that extended structures are not a universal property of complex datasets.  Nevertheless, even this relatively small dataset had no exploitable separations in feature space, but did exhibit density variations that could be used to obtain significant results.

Although the mathematics behind DQC is not familiar outside of the physics community, one doesn't need to understand it to use the DQC software.  A DQC analysis is simple to carry out and usually requires only a few commands.  The visual nature of the output creates the impression of being immersed in the data.  The movie that results from DQC evolution presents a powerful visual record of the entire process and permits one to better understand what is happening.  Of course, once interesting structures reveal themselves, standard statistical methods can be used to establish the significance of the results.  An important next step in the development of DQC will be learning how to best combine this technique with standard statistical tools in order to maximize the power of both methodologies.  An important new capability that DQC brings to the table is that - by using DQC - data producers can embark upon an initial exploration of unstructured data without having to wait for skilled data analysis resources to become available.

**References:**

1. M. Weinstein and D. Horn, Phys. Rev. E, **80**, 066117 (2009).
2. D. Horn and A. Gottlieb, Phys. Rev. Lett., **88**, 018702 (2001).
3. Parzen, E., Annals of Mathematical Statistics, **33,** 1065–1076, (1962).
4. R. O. Duda, P. E. Hart, and D. G. Stork, Pattern Classification, 2nd ed. Wiley-Interscience, New York, (2001).
5. J.C. Andrews, F. Meirer, Y. Liu, Z. Mester, P. Pianetta, *Transmission X-Ray Microscopy for Full-Field Nano Imaging of Biomaterials*, MICROSCOPY RESEARCH AND TECHNIQUE, 2011 Jul;**74(7)**:671-81. doi: 10.1002/jemt.20907.
6. F. Meirer, J. Cabana, Y. Liu, A. Mehta, J. C. Andrews and P. Pianetta, J. Synchrotron Radiat. **18**, 773 (2011); Y. Liu, F. Meirer, P.A. Williams, J. Wang, J.C. Andrews, and P. Pianetta, *TXM-Wizard: a program for advanced data collection and evaluation in full-field transmission X-ray microscopy*, Journal of Synchrotron Rad. (2012). **19**, 281–287; F. Meirer, Y. Liu, E. Pouyet, B. Fayard, M. Cotte, C. Sanchez, J.C. Andrews, A. Mehta, and P. Sciau, *Probing the State of Technological Sophistication of an Ancient Civilization: Estimating Firing Conditions for Roman Ceramics*, Journal of Analytical Atomic Spectrometry (2013), submitted

7.  Ph. Sciau, Y. Leon, Ph. Goudeau, S.C. Sirine, S. Webb, A. Mehta, J. Anal. At. Spectrom. **26** (5), 969-976, (2011). (on line, DOI:10.1039/C0JA00212G).
8.  The data for this analysis was provided to us by David Reis and Mariano Trigo in the department of Pulse Research at the SLAC National Accelerator Laboratory
9.  The data for this analysis was provided to us by Haibin Su and Lin Xin at the NTU- National Technical University, Singapore.
10. The DQC analysis of the aquaporin data was done by Marvin Weinstein, the correlated sequence analysis (CSA) was carried out by Lin-Xin at the NTU- National Technical University, Singapore.
11. Shapira, A., Hofstetter, A. 1993. Source parameters and scaling relationships of earthquakes in Israel. Tectonophysics **217**: 217-226.
12. Baer, G., Funning, G.J., Shamir, G., and Wright, T.J., 2008. The 1995 November 22, Mw=7.2 Gulf of Elat earthquake cycle revisited. *Geophys. J. Int.* **175**, 1040–1054.
13. G. Shaked, M.Sc. thesis, available at http://horn.tau.ac.il/publications/Thesis_GS.pdf.
14. Wessel, P., and Smith, W. 1991. Free software helps maps and display data. EOS Trans. AGU 72, 441.
15. Hofstetter, A., Thio, H.-K., and Shamir, G., 2003. Source mechanism of the 22/11/1995 Gulf of Aqaba earthquake and its aftershock sequence. *Journal of Seismology*, **7**, 99-114.
16. Hofstetter, A., 2003. Seismic observations of the 22/11/1995 Gulf of Aqaba earthquake sequence. Tectonophysics, **369**, 21-36.
17. Michael C. Munnix, Takashi Shimada, Rudi Schafer, Francois Leyvraz, Thomas H. Seligman, Thomas Guhr, and H. Eugene Stanley, 2012. Identifying States of a Financial Market. Scientific Reports 2, 644. doi:10.1038/srep00644

**Acknowledgments:**

The authors would like to thank Burton Richter, David Hitlin, Robert Cahn, James D. Bjorken and Alfred S. Goldhaber for helpful remarks. Data used in this publication was collected at the Stanford Synchrotron Radiation Lightsource, a Directorate of SLAC National Accelerator Laboratory and an Office of Science User Facility operated for the U.S. Department of Energy Office of Science by Stanford University. The earthquake analysis was partially supported by the Earth Sciences and Research Administration, Ministry of Energy, Israel.

# Appendix A. Initial DQC Evolution of LCLS Pump-probe Data

While plotting the four SVD components shown in Fig.9 strongly suggests that we should focus attention on the data points that lie in the broadened Bragg peaks, we nevertheless carried out

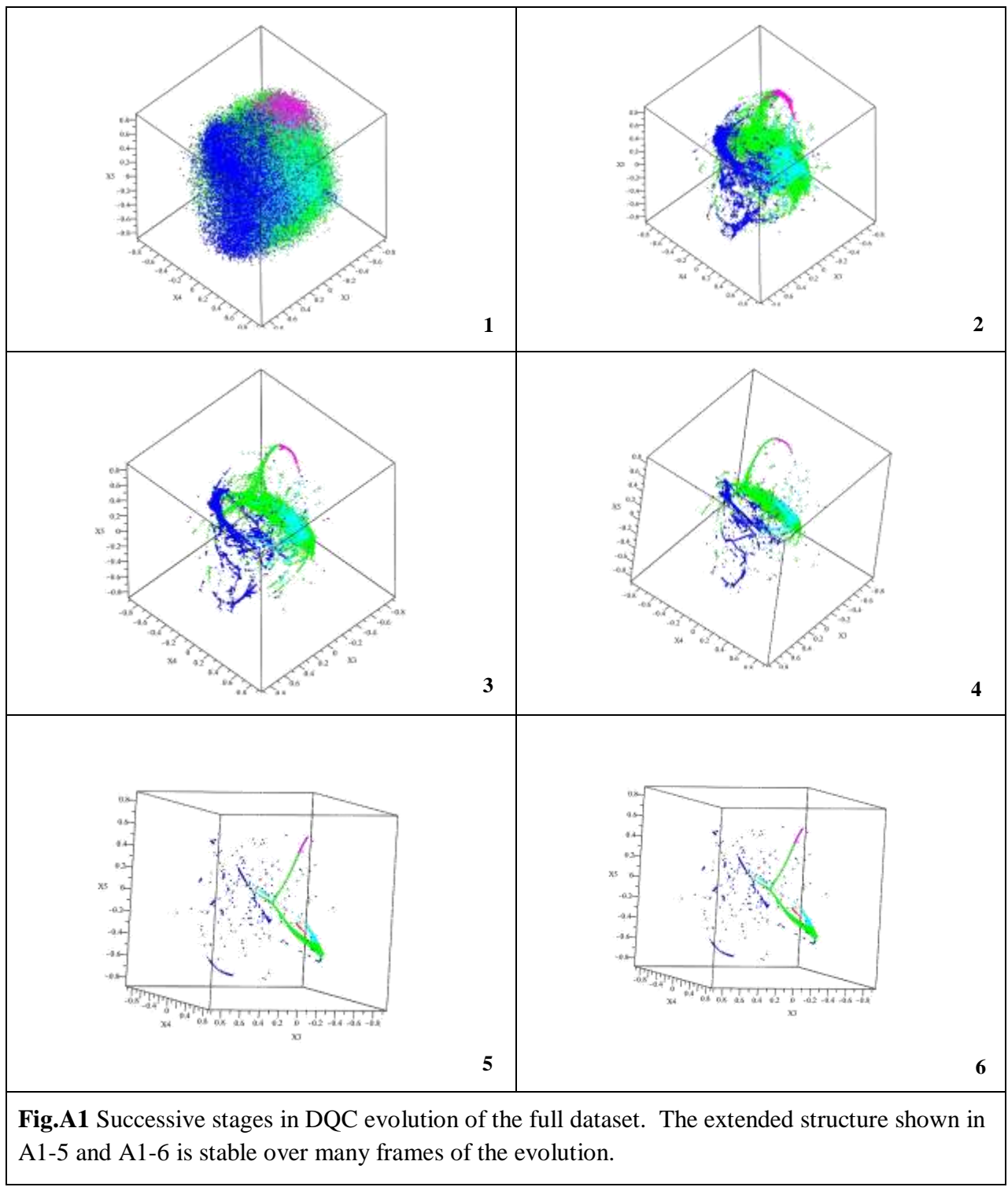

**Fig.A1** Successive stages in DQC evolution of the full dataset. The extended structure shown in A1-5 and A1-6 is stable over many frames of the evolution.

DQC evolution for all of the data in order to see if it contained any surprises. To remove noise – as with the TXM-XANES data - we worked with the data as reconstructed by the first 9 SVD components. Fig.A1 shows successive frames of the animation produced by DQC evolution. From this we see that the final result is many point-like clusters, a few short extended structures and one long extended structure reminiscent of the with pattern of amplitude variation shown in the background noise as seen in Fig.9. To understand the extended structure we color the four distinct parts of this long extended structure magenta, green, cyan and red. Averaging the spectra for the points contained in each of these components separately, we produce the four curves shown in Fig.A2-1. Clearly these curves look very similar except for amplitude and – as is shown in Fig.A2-2 – if we normalize these curves all to be unity at the endpoint, then they almost exactly coincide with one another. In the end, more careful analysis shows that all of the extended structures coincide with the background noise in the detector (*i.e.,* the region outside of the broadened Bragg peaks) and in all they constitute over 95% of the data. The amplitude variation of the signal in these spectra is consistent with background regions. Thus, this analysis supported the decision to select only the data corresponding to the two broadened Bragg peaks. This cut reduced the data to ~15,000 data points. This smaller dataset was once again analyzed using DQC evolution in order to isolate the good data and find correlated pixels.

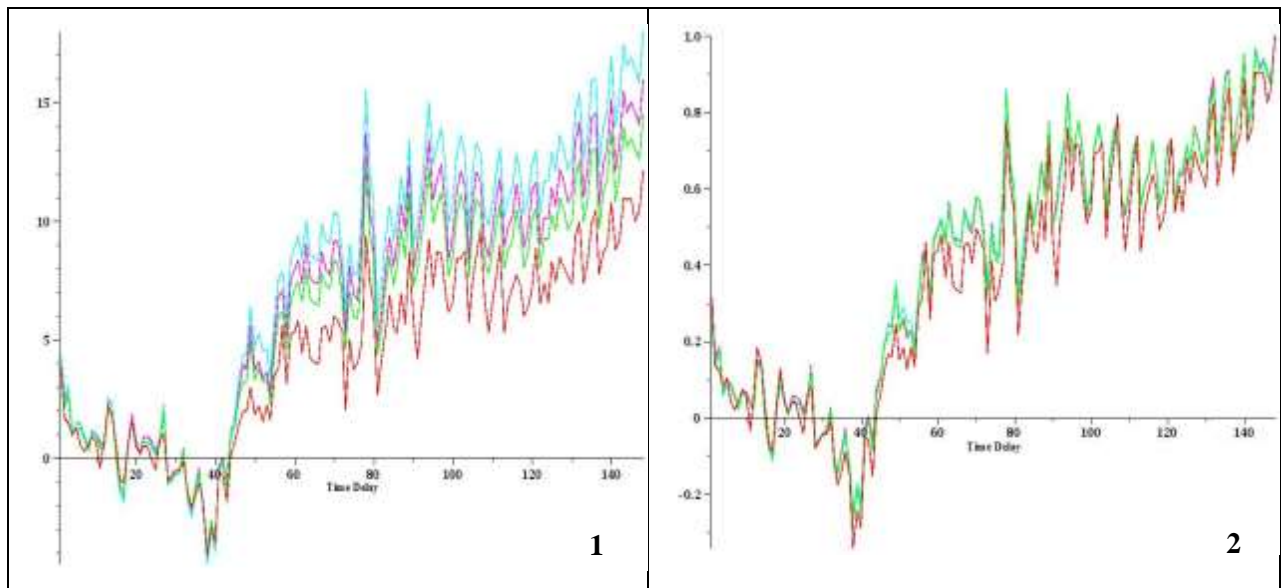

**Fig.A2 (1)** Plots of different segments of the extended structure appearing after DQC evolution showing their similarity to one another. (2) The same curves after normalizing the last point in each curve to have the value 1.

**Appendix B. The Second DQC Evolution of filtered LCLS Pump-probe Data**

Fig.B1 shows four frames from the quantum evolution for the filtered data of the reduced dataset. The first row are four frames coming from dimensions 123, the second row are four frames from the 456 and 789 respectively. It is important to observe that most points coalesce into compact clusters very quickly, but certain subsets of the data persist as extended structures for quite some time - before eventually collapsing to a point. As in the case of the TXM-XANES data the transient structures indicate clusters that can be subdivided to obtain spectra that differ slightly from one another in ways that can be simply parameterized.

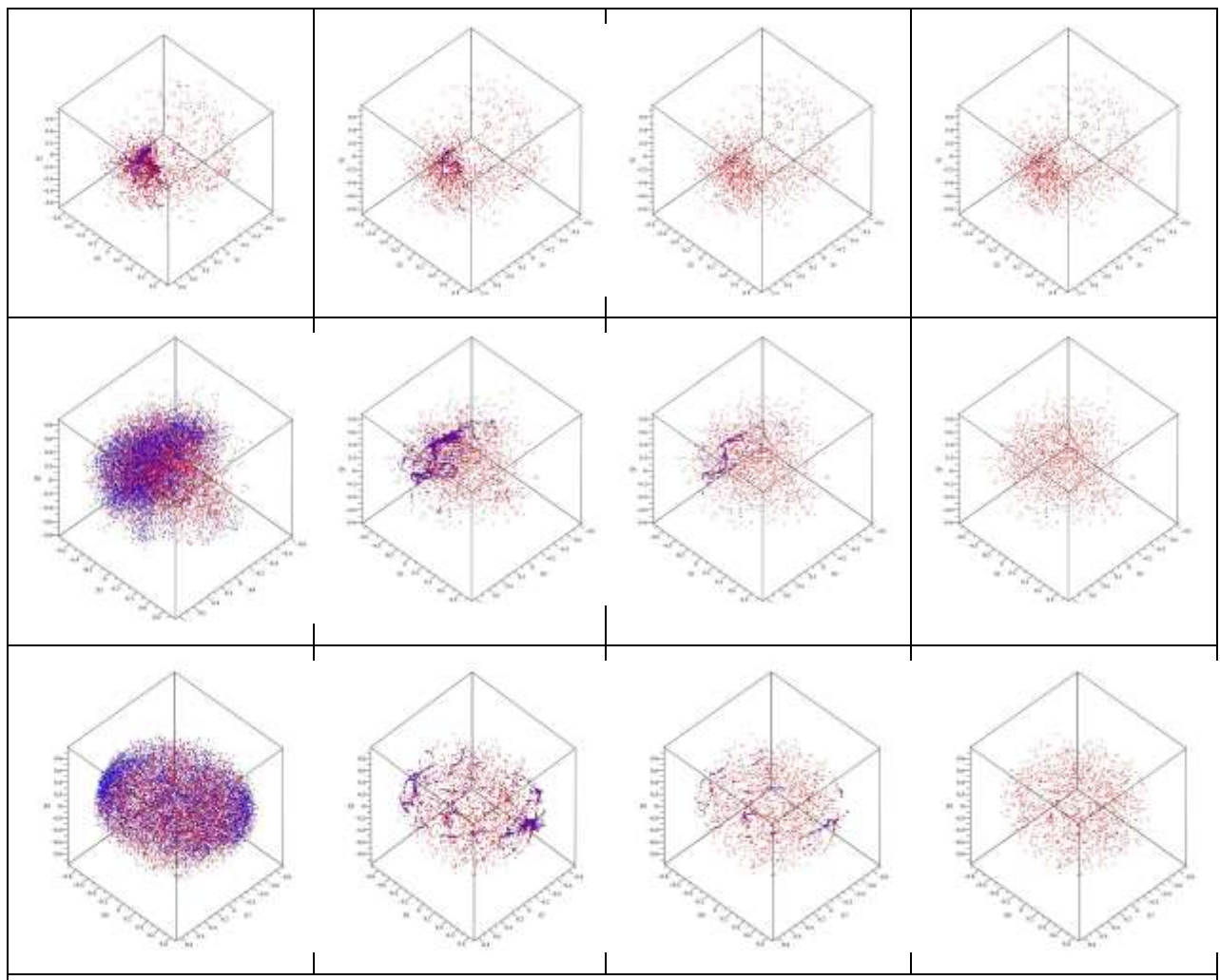

**Fig.B** Evolution of th 9-dimensional data over hundreds of frames. The first row shows the evolution as it appear in dimensions 1,2,3. The second row shows dimensions 4,5,6 and the last row shows dimension 7,8,9. Note that most point-like clusters appear early in the evolutionary process, however some transient extended structures appear and persist for well over 100 frames. Comparison of spectra in the extended clusters show small but significant variation from one another.

37